\documentclass[reprint,superscriptaddress,showpacs,preprintnumbers,amsmath,amssymb,aps,prb]{revtex4-1}
\usepackage{graphicx}
\usepackage{dcolumn}
\usepackage{bm}
\usepackage{multirow}
\usepackage{color}
\usepackage{ulem}

\newcommand{\1}{\phantom{0}}

\begin{document}

\title{Band widths and gaps from the Tran-Blaha functional : Comparison with many-body perturbation theory.}

\author{David Waroquiers}
\email{david.waroquiers@uclouvain.be}
\affiliation{Universit\'e catholique de Louvain, Institute of Condensed Matter and Nanosciences, NAPS Chemin des Etoiles 8, bte L7.03.01, B-1348 Louvain-la-Neuve, Belgium}
\affiliation{European Theoretical Spectroscopy Facility}
\author{Aur\'elien Lherbier}
\affiliation{Universit\'e catholique de Louvain, Institute of Condensed Matter and Nanosciences, NAPS Chemin des Etoiles 8, bte L7.03.01, B-1348 Louvain-la-Neuve, Belgium}
\affiliation{European Theoretical Spectroscopy Facility}
\author{Anna Miglio}
\affiliation{Universit\'e catholique de Louvain, Institute of Condensed Matter and Nanosciences, NAPS Chemin des Etoiles 8, bte L7.03.01, B-1348 Louvain-la-Neuve, Belgium}
\affiliation{European Theoretical Spectroscopy Facility}
\author{Martin Stankovski}
\affiliation{Universit\'e catholique de Louvain, Institute of Condensed Matter and Nanosciences, NAPS Chemin des Etoiles 8, bte L7.03.01, B-1348 Louvain-la-Neuve, Belgium}
\affiliation{European Theoretical Spectroscopy Facility}
\author{Samuel Ponc\'e}
\affiliation{Universit\'e catholique de Louvain, Institute of Condensed Matter and Nanosciences, NAPS Chemin des Etoiles 8, bte L7.03.01, B-1348 Louvain-la-Neuve, Belgium}
\affiliation{European Theoretical Spectroscopy Facility}
\author{Micael J. T. Oliveira}
\affiliation{Center for Computational Physics, University of Coimbra, Portugal}
\affiliation{European Theoretical Spectroscopy Facility}
\author{Matteo Giantomassi}
\affiliation{Universit\'e catholique de Louvain, Institute of Condensed Matter and Nanosciences, NAPS Chemin des Etoiles 8, bte L7.03.01, B-1348 Louvain-la-Neuve, Belgium}
\affiliation{European Theoretical Spectroscopy Facility}
\author{Gian-Marco Rignanese}
\affiliation{Universit\'e catholique de Louvain, Institute of Condensed Matter and Nanosciences, NAPS Chemin des Etoiles 8, bte L7.03.01, B-1348 Louvain-la-Neuve, Belgium}
\affiliation{European Theoretical Spectroscopy Facility}
\author{Xavier Gonze}
\affiliation{Universit\'e catholique de Louvain, Institute of Condensed Matter and Nanosciences, NAPS Chemin des Etoiles 8, bte L7.03.01, B-1348 Louvain-la-Neuve, Belgium}
\affiliation{European Theoretical Spectroscopy Facility}
\date{\today}

\begin{abstract}
For a set of ten crystalline materials (oxides and semiconductors), we compute the electronic band structures using the Tran-Blaha [Phys. Rev. Lett. {\bf 102}, 226401 (2009)] (TB09) functional. The band widths and gaps are compared with those from the local-density approximation (LDA) functional, many-body perturbation theory (MBPT), and experiments. At the density-functional theory (DFT) level, TB09 leads to band gaps in much better agreement with experiments than LDA. However, we observe that it globally underestimates, often strongly, the valence (and conduction) band widths (more than LDA). MBPT corrections are calculated starting from both LDA and TB09 eigenenergies and wavefunctions. They lead to a much better agreement with experimental data for band widths. The band gaps obtained starting from TB09 are close to those from quasi-particle self-consistent $GW$ calculations, at a much reduced cost. Finally, we explore the possibility to tune one of the semi-empirical parameters of the TB09 functional in order to obtain simultaneously better band gaps and widths. We find that these requirements are conflicting.
\end{abstract}


\maketitle

\section{\label{sec:introduction}Introduction}
In the last thirty years, Density Functional Theory~\cite{Hohenberg1964,Kohn1965} (DFT) has been successfully applied to the computation of structural, mechanical and linear-response properties of many different systems.~\cite{Martin2004,Burke2012} One of the major shortcomings of DFT is related to the prediction of electronic band gaps, known as \textit{band gap problem}.~\cite{Sham1985} Because DFT is a ground-state theory, Kohn-Sham eigenvalues cannot formally be interpreted as quasi-particle ones. In order to obtain reliable band gaps, one should not describe the electronic interactions using a mean-field approach such as DFT. On the contrary, this is appropriately tackled by using Many-Body Perturbation Theory (MBPT), as introduced by Hedin and Lundqvist.~\cite{Hedin1965,Hedin1970} MBPT is computationally much more expensive than DFT, and some approximations have to be made for real applications. Up to now, most of the studies of real systems conducted within MBPT have used the so-called one-shot $GW$ approximation with a long-standing record of success.~\cite{Aulbur2000}

However, even for the one-shot $GW$ method, the computational cost is about two orders of magnitude higher than standard DFT. Many attempts have therefore been made to obtain better band gaps and to improve the quality of the excited states without using any computationally demanding many-body techniques. The $\Delta$SCF method~\cite{Martin2004} for atoms and molecules, further generalized to solids by Chan and Ceder as $\Delta$-sol,~\cite{Chan2010} leads to fundamental band gaps in closer agreement with experiment. However, the full band structure of the excited states remains quite approximate. Hybrid functionals can have a positive effect on the accuracy of the band gap and the position of higher excited states~\cite{Marques2011} but are more computationally demanding than standard exchange and correlation (XC) functionals.

In a recent letter by Tran and Blaha,~\cite{Tran2009} a modified version of the Becke-Johnson exchange potential~\cite{Becke2006} combined with an LDA correlation part was proposed to improve band gap predictions in DFT. This new XC functional (referred to as TB09) has already been applied to a large variety of systems~\cite{Singh2010,Dixit2012,Hetaba2012,Feng2010, Smith2012, Kresse2012,Camargo2012} and was able to predict band gaps in close agreement with experiment. However, limitations of this functional have been pointed out by some authors. First, there is no energy functional from which the potential is derived,~\cite{Karolewski2009,Gaiduk2009,Camargo2012} which prevents the description of any energy-related properties, such as defect formation energies or phase stability. Secondly, it is not intended to be used for systems without an electronic gap and indeed seems to be less accurate for such systems.~\cite{Koller2011} Finally, effective masses are reported to be overestimated,~\cite{Kim2010, Kresse2012} with an accompanying narrowing of the bands. There are other known drawbacks to the TB09 functional, namely the fact that it is not size consistent and that it is not gauge invariant.~\cite{Raesaenen2010}

In this work, a systematic comparison of electronic band structures obtained from the LDA and TB09 functionals is performed. These are compared with the more accurate band structures obtained through perturbative $GW$ calculations, and available experimental data. Several oxides and other technologically or theoretically interesting materials are investigated : silicon, germanium, silicon dioxide, tin monoxide and dioxide, zinc oxide, calcium sulfide, calcium oxide, magnesium oxide and lithium fluoride. We find that TB09 indeed delivers band gap in much better agreement with perturbative $GW$ calculations and experiment than the simple LDA. By contrast, valence band widths are contracted from LDA to TB09 by a narrowing factor varying between 1.0 and 1.6, and often strongly disagree with both perturbative $GW$ calculations and experiment.\cite{xps_broadening} Our perturbative $GW$ calculations are performed starting from LDA eigenvalues and wavefunctions (these will be referred to as $G_0W_0$@LDA), but also from TB09 eigenvalues and wavefunctions (referred to as $G_0W_0$@TB09). The difference is non-negligible. In fact, the one-shot $GW$ band gaps obtained starting from TB09 are close to those from quasi-particle self-consistent $GW$ calculations, at a much reduced cost.
In order to produce perturbative $GW$ band structure, we introduce a new approximate interpolation scheme referred to as \textit{polyfit}. It is based on a piecewise polynomial approach, and is especially simple to implement.

In Sec.\ \ref{sec:methods}, the basic theories used throughout this study are reviewed. The main ingredients of DFT are first introduced as well as the approximations to which they are intrinsically related. The form of the LDA and TB09 XC functionals is also specified. Then, implications of Hedin's equations and the Many-Body problem are summarized. In particular, the $GW$ method and the associated technical details are surveyed. Finally, the \textit{polyfit} interpolation scheme is described. In Sec.\ \ref{sec:results}, the band structures, densities of states, band gaps and band widths obtained with LDA and TB09 using DFT and $GW$ are presented for the ten above-mentioned materials. Then, in Sec.\ \ref{sec:discussion}, the results and consequences of the use of the TB09 XC functional in DFT-based electronic predictions are discussed. In particular, we consider the possibility to tune the semi-empirical parameter appearing in the TB09 in order to achieve better band widths. However, this tuning of the band width cannot be achieved without degrading the description of the band gaps. The final conclusions are drawn in Sec. \ref{sec:conclusion}.

\section{\label{sec:methods}Methods}

All the calculations were done with the ABINIT software package~\cite{Gonze2005, Gonze2009} using norm-conserving pseudopotentials. These were prepared using the Atomic Pseudopotential Engine.~\cite{Oliveira2008} The \textsc{Libxc}, a library of XC functionals,~\cite{Marques2012} has been utilized for the evaluation of the XC terms. Hereafter, the basic theories, methods and approximations that are used throughout this study are presented.

\subsection{\label{subsec:dft}Density functional theory}

In DFT, the ground-state of a system is obtained by solving a set of self-consistent equations (\ref{eqn:KS_scf_equations_oneelectronschrod}-\ref{eqn:KS_scf_equations_kspotential}), as proposed by Kohn and Sham.~\cite{Kohn1965} The wavefunctions and energies follow from the solution of the one-electron Schr\"odinger equation :

\begin{equation}
\left(-\frac{1}{2}\nabla^2 + V_{\mathrm{KS}}(\mathbf{r})\right)\psi_i^{\mathrm{KS}}(\mathbf{r}) = \epsilon_i^{\mathrm{KS}} \psi_i^{\mathrm{KS}}(\mathbf{r}), \label{eqn:KS_scf_equations_oneelectronschrod}
\end{equation}

\noindent where the Kohn-Sham potential $V_{\mathrm{KS}}(\mathbf{r})$ relies on the external potential $V_{\mathrm{ext}}(\mathbf{r})$ resulting from the electrostatic potential of the ions as well as the electrostatic potential $V_{\mathrm{H}}(\mathbf{r})$ originating from the electronic density $n(\mathbf{r})$

\begin{equation}
V_{\mathrm{KS}}(\mathbf{r}) = V_{\mathrm{ext}}(\mathbf{r})+V_{\mathrm{H}}(\mathbf{r})+V_{\mathrm{xc}}[n(\mathbf{r})]. \label{eqn:KS_scf_equations_kspotential}
\end{equation}

Therefore, the unique approximation is the XC potential, which is usually expressed as a functional derivative of the XC energy : $V_{\mathrm{xc}}[n(\mathbf{r})]$=$\frac{\delta E_{\mathrm{xc}}\left[n\right]}{\delta n(\mathbf{r})}$. If the exact XC functional were known, the method would in principle be exact. In practice, only approximations of this functional are available for non-trivial systems. There are many proposed approximations in the literature and the simplest approach is to use the local density approximation~\cite{Hohenberg1964,Kohn1965} for the XC energy $E_{\mathrm{xc}}[n]$ which is expressed as

\begin{equation}\label{eqn:Exc_LDA}
E^{\mathrm{LDA}}_{\mathrm{xc}}[n]=\int \mathrm{d}\mathbf{r}\,n(\mathbf{r})\epsilon_{\mathrm{xc}}(n),
\end{equation}

\noindent with $\epsilon_{\mathrm{xc}}(n)$ being XC energy density of the homogeneous electron gas of density $n$. In this work, the Perdew-Wang~\cite{Perdew1992a} parametrization for the LDA XC energy, based on quantum Monte Carlo data from Ceperley and Alder~\cite{Ceperley1980} is used.

The XC potential proposed by Tran and Blaha~\cite{Tran2009} is composed of a modified version of the Becke-Johnson~\cite{Becke2006} exchange potential and a LDA correlation part. The exchange part $V^{\mathrm{TB09}}_{\mathrm{x}}$ takes the following form :

\begin{equation}\label{eqn:vx_tb09}
V^{\mathrm{TB09}}_{\mathrm{x}}=cV_{\mathrm{x}}^{\mathrm{BR}}(\mathbf{r})+(3c-2)\frac{1}{\pi}\sqrt{\frac{5}{12}}\sqrt{\frac{2t_s(\mathbf{r})}{n(\mathbf{r})}},
\end{equation}

\noindent where $V_{\mathrm{x}}^{\mathrm{BR}}(\mathbf{r})$ is the Becke-Roussel potential~\cite{Becke1989} modeling the Coulomb potential created by the exchange hole and $t_s(\mathbf{r})$=$\frac{1}{2}\sum_{i=1}^{N}\nabla \psi_i^\ast \cdot \nabla \psi_i$ is the kinetic energy density. The parameter $c$ results from the following equation :

\begin{equation}
c=\alpha+\beta\left(\frac{1}{V_{\mathrm{cell}}}\int_{\mathrm{cell}}d\mathbf{r'}\frac{|\nabla n(\mathbf{r'})|}{n(\mathbf{r'})}\right)^{1/2},
\end{equation}

\noindent where $V_{\mathrm{cell}}$ is the unit cell volume and where $\alpha$ and $\beta$ were fitted to experimental gaps in the original paper using a least-square procedure.

In this study, the DFT electronic structures obtained using XC potentials from both the LDA and TB09 functionals are compared.
These two sets of KS states are then used as initial starting points for perturbative $GW$ corrections ($G_0W_0$@LDA and $G_0W_0$@TB09).

\subsection{\label{subsec:mbpt}Many-body perturbation theory and the $GW$ approximation}

In the Many-body perturbation theory (MBPT), the energies and wavefunctions of the system are obtained by solving the quasi-particle (QP) equation :

\begin{multline}
\left(-\frac{1}{2}\nabla^2 + V_{\mathrm{ext}}(\mathbf{r})+V_{\mathrm{H}}(\mathbf{r})\right)\psi_{i}^{\mathrm{QP}}(\mathbf{r}) \\
+ \int d\mathbf{r'} \Sigma(\mathbf{r},\mathbf{r'};\omega=\epsilon_{i}^{\mathrm{QP}})\psi_{i}^{\tiny{\mathrm{QP}}}(\mathbf{r}) = \epsilon_{i}^{\mathrm{QP}} \psi_{i}^{\mathrm{QP}}(\mathbf{r}), \label{eqn:quasiparticle_equation}
\end{multline}

\begin{figure}
\center
\includegraphics{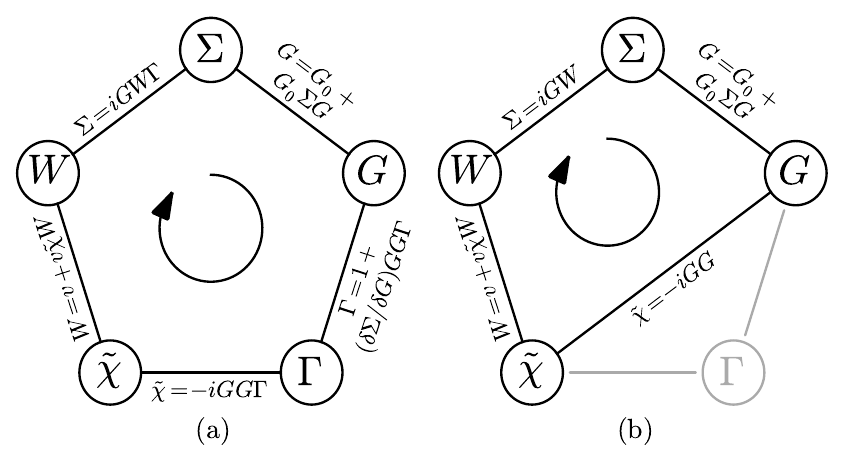}
\caption{\label{fig:mpbt-gw-pentagons}{(a) Schematic representation Hedin's equations and (b) $GW$ approximation.}}
\end{figure}

This equation is very similar to Eq. (\ref{eqn:KS_scf_equations_oneelectronschrod}) except that the self-energy $\Sigma$ (a non-local, energy-dependent and non-hermitian operator) takes the role of the XC potential $V_{\mathrm{xc}}$. In a completely general theory, the self-energy is obtained by solving a set of five self-consistent integro-differential equations which were originally introduced by Hedin and Lundqvist~\cite{Hedin1965,Hedin1970} as illustrated schematically in Fig.~\ref{fig:mpbt-gw-pentagons}(a). In practice, the complexity of the problem is often reduced by assuming a unitary vertex operator $\Gamma$=$1$ leading to the so-called $GW$ approximation as depicted in Fig.~\ref{fig:mpbt-gw-pentagons}(b). The self-energy $\Sigma$ is represented as the product of the Green's function $G$ and the screened Coulomb potential $W$ :

\begin{equation}\label{eqn:sigma_equation}
\Sigma(\mathbf{r},\mathbf{r'};\omega)=\frac{i}{2\pi}\int d\omega' G(\mathbf{r},\mathbf{r'},\omega+\omega')W(\mathbf{r},\mathbf{r'},\omega')
\end{equation}

It is common practice to avoid the full self-consistency by stopping after the first iteration. This has been shown to be a good approximation, provided that the initial (KS) wavefunctions and eigenvalues are close to the QP ones. The resulting method which is known as the \textit{one-shot} $GW$ or $G_0W_0$ approximation uses a perturbative approach for the evaluation of the QP corrections to the KS eigenvalues :

\begin{equation*}
\epsilon_i^{QP}=\epsilon_i^{KS}+\Delta\epsilon^{GW}_i
\end{equation*}
\noindent with
\begin{multline}\label{eqn:perturbative_g0w0}
\Delta\epsilon^{GW}_i= \\
Z_i\left\langle \psi_i^{\mathrm{KS}}|\Sigma(\mathbf{r},\mathbf{r'};\epsilon_i^{\mathrm{KS}})-V_{\mathrm{xc}}(\mathbf{r})\delta(\mathbf{r}-\mathbf{r'})|\psi_i^{\mathrm{KS}} \right\rangle
\end{multline}
\noindent where $Z_i^{-1}$=$1-\left\langle\psi_i^{\mathrm{KS}}\right|\left. \frac{\partial\Sigma(\omega)}{\partial \omega}\right|_{\omega=\epsilon_i^{\mathrm{KS}}}\left| \psi_i^{\mathrm{KS}} \right\rangle$ is the renormalization factor and $i$ stands for the spin, \textit{k}-point and band indexes.

It is important to stress here that the initial KS starting point can have a strong influence on the one-shot $GW$ electronic structure obtained and in particular, the band gaps. For what concerns this study, we highlight that the wavefunctions $\psi_i^{\mathrm{KS}}$ and eigenvalues $\epsilon_i^{\mathrm{KS}}$ which appear in Eq.~(\ref{eqn:perturbative_g0w0}) strongly depend on the XC functional.

The Godby and Needs~\cite{Godby1989} plasmon-pole model has been used to account for the frequency dependence of the screening in Eq.~(\ref{eqn:sigma_equation}). This model has been demonstrated to accurately reproduce the states within a large range of energies around the Fermi level.~\cite{Stankovski2011,Miglio2012} Unless explicitly mentioned, the extrapolar method~\cite{Bruneval2008} with a compensation energy of 2 Ha has been used in order to reduce the number of bands included in the computation of the $GW$ corrections.

For eight out of the ten above-mentioned materials that we study, we also provide band gaps obtained from the Quasi-particle Self-consistent $GW$ (QS$GW$) approximation, introduced by van Schilfgaarde \textit{et al.}~\cite{Schilfgaarde2006} Such QS$GW$ band gaps do not depend on the starting point and tend to slightly overestimate the experimental band gap value.

\subsection{\label{subsec:polyfit}Interpolation of $GW$ eigenvalues : the \textit{polyfit} energy approximation}

The representation of a sufficiently accurate band structure (not only the band gap) or density of states requires a dense sampling of \textit{k}-points in the Brillouin zone. It is preferable to avoid such calculations by interpolating the $GW$ eigenvalues on a very dense mesh from the results obtained on a coarser \textit{k}-point grid.

One possible technique relies on maximally-localized Wannier functions (MLWFs).~\cite{Marzari1997,Souza2001} The production of $GW$ band structure using Wannier functions is described e.g. in Ref.~\onlinecite{Shaltaf2009}. For valence states, this technique easily applies. Indeed, the valence wavefunctions are well localized and the generation of the MLWFs is almost instantaneous. On the other hand, applying this procedure for the conduction states is not an easy task as these states are usually more delocalized : finding proper Wannier functions is far from being automatic. For this reason a different and much more straightforward procedure is adopted here.

It is based on the fact that, at least for simple materials such as the ones examined here, the $GW$ correction to DFT eigenvalues is, to a large extent, only a function of the energy. Fluctuations, related to wavevector and state dependences, are present, but can be ignored when the computation of the band structure is done for plotting purposes. For more complex materials, the $GW$ corrections are no longer a simple function of the DFT eigenvalues, as observed e.g. in Ref.~\onlinecite{Vidal2010a}, and the method used here may not be sufficient.

In our \textit{polyfit} approach, the $GW$ corrections to eigenvalues are piecewise interpolated using several least-squares polynomial fits corresponding to different energy ranges. In this way, the dependence of the $GW$ correction $\Delta\epsilon^{\mathrm{GW}}_i$ on the band index and \textit{k}-point $i$=$(n,\mathbf{k})$ is replaced by a function of the KS eigenvalues only :

\begin{equation}
\Delta\epsilon^{\mathrm{GW}}_i\longrightarrow\Delta E^{\mathrm{GW}}(\epsilon_i^{\mathrm{KS}})
\end{equation}

First, the electronic bands are split into separate groups depending on their energy range, using midpoints in forbidden energy gap as limits of the different intervals (so-called \textit{energy pivots}). For each interval, a third-order polynomial is employed as a fit for the $GW$ corrections :

\begin{equation*}
\Delta E^{\mathrm{GW}}(\epsilon_i^{\mathrm{KS}})=a_0+a_1\epsilon_i^{\mathrm{KS}}+a_2(\epsilon_i^{\mathrm{KS}})^2+a_3(\epsilon_i^{\mathrm{KS}})^3
\end{equation*}

\begin{figure}
\center
\includegraphics{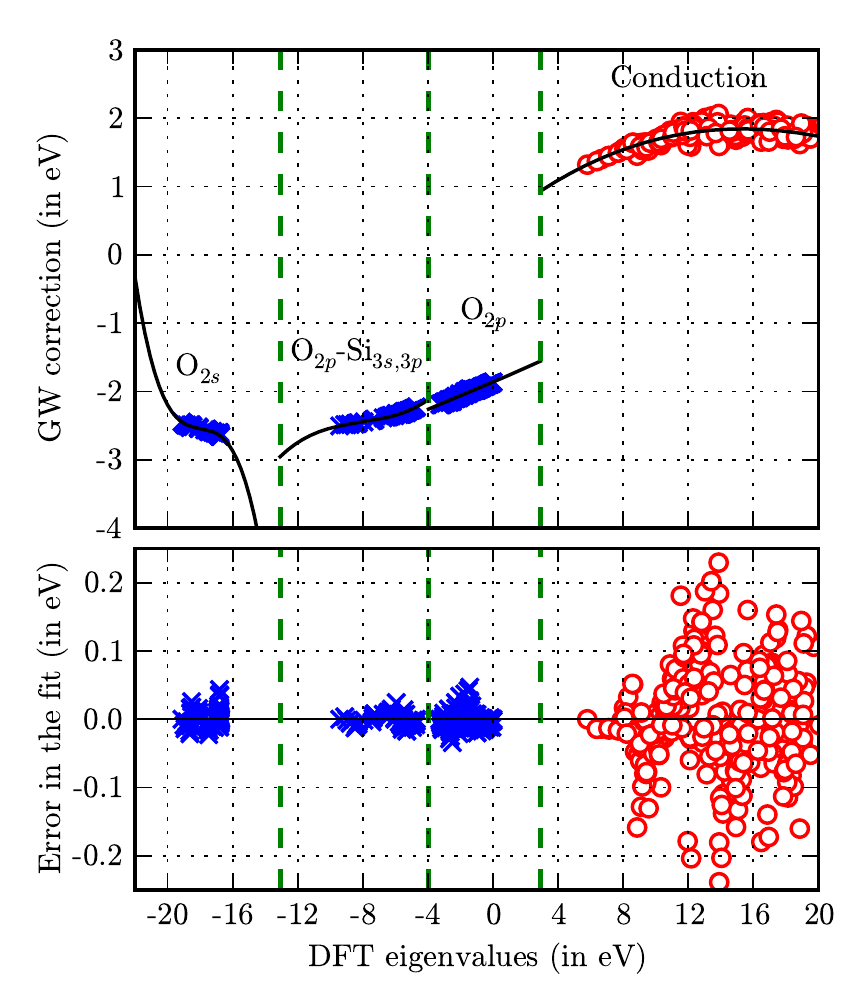}
\caption{\label{fig:polyfit}{Interpolation of $GW$ eigenvalues with the \textit{polyfit} energy approximation. Example for $\alpha$-quartz in LDA. Upper panel : $G_0W_0$ corrections for the valence (blue crosses) and conduction (red circles) states with respect to DFT eigenvalues. The vertical dashed green lines corresponds to the \textit{energy pivots} chosen to separate the different groups of bands. The black lines are the polynomial fits obtained for each group of band. Lower panel : Estimation of the error made by the \textit{polyfit} approximation used in the upper panel.}}
\end{figure}

The polynomial is constrained to go through the endpoints of each group of bands in order to preserve the band widths. Additionally, the spread of the coarse mesh eigenvalues with respect to the polynomial is minimized by adjusting the polynomial coefficients. Finally, the interpolation polynomials are used to correct the DFT band structure and DOS on a fine mesh. An example of this fitting procedure is given in Fig.~\ref{fig:polyfit} for the case of $\alpha$-quartz. The upper panel shows the polynomial fits obtained for the $GW$ corrections.

Such an interpolation procedure is obviously not perfect. The error $\theta_i$ made for each state $i$ belonging to the coarse grid when using this interpolation procedure gives an estimate of the overall error of the fit. For each explicitly calculated state on the coarse grid, it is evaluated as the difference between the \textit{polyfit} and the real correction $\Delta\epsilon^{\mathrm{GW}}_i$ from Eq.~(\ref{eqn:perturbative_g0w0}) :

\begin{equation*}
\theta_i=\Delta E^{\mathrm{GW}}(\epsilon_i^{\mathrm{KS}})-\Delta\epsilon^{\mathrm{GW}}_i
\end{equation*}

The errors for $\alpha$-quartz are given in the lower panel of Fig.~\ref{fig:polyfit}. For all the considered valence states and for the lowest conduction states (up to $3$ eV above the conduction band minimum), the error is always smaller than $0.1$ eV whereas for higher-energy states displayed in the figure, the error goes up to $0.25$ eV. For all the systems investigated in the present study, the error is less than $0.2$ eV for the valence states and the lowest conduction states, and less than $0.5$ eV for higher-energy states.

\section{\label{sec:results}Results}

\subsection{Silicon and germanium}

The experimental silicon lattice parameter~\cite{Wyckoff1960} $a$=$5.43$~\AA~was used in this study. Silicon has an indirect band gap from $\Gamma$ to a point located about $80\%$ of the way along the path $\Gamma$-$X$. Germanium possesses the same  zinc-blende crystalline structure ($Fd\overline{3}m$) with an experimental lattice parameter~\cite{Wyckoff1960} $a$=$5.66$~\AA~and has an indirect band gap from $\Gamma$ to $L$.

The wavefunctions have been expanded in a plane wave basis set with a kinetic energy cutoff of $20$ Ha. The \textit{k}-point grids used were 6$\times$6$\times$6 unshifted and 8$\times$8$\times$8 unshifted meshes for silicon (Si) and germanium (Ge) respectively. For $GW$ calculations, the size of the dielectric matrix used was determined by a kinetic energy cutoff of $10$ Ha for Si and $20$ Ha for Ge, and 200 bands have been used.

\begin{figure*}[ht!]
\center
\includegraphics{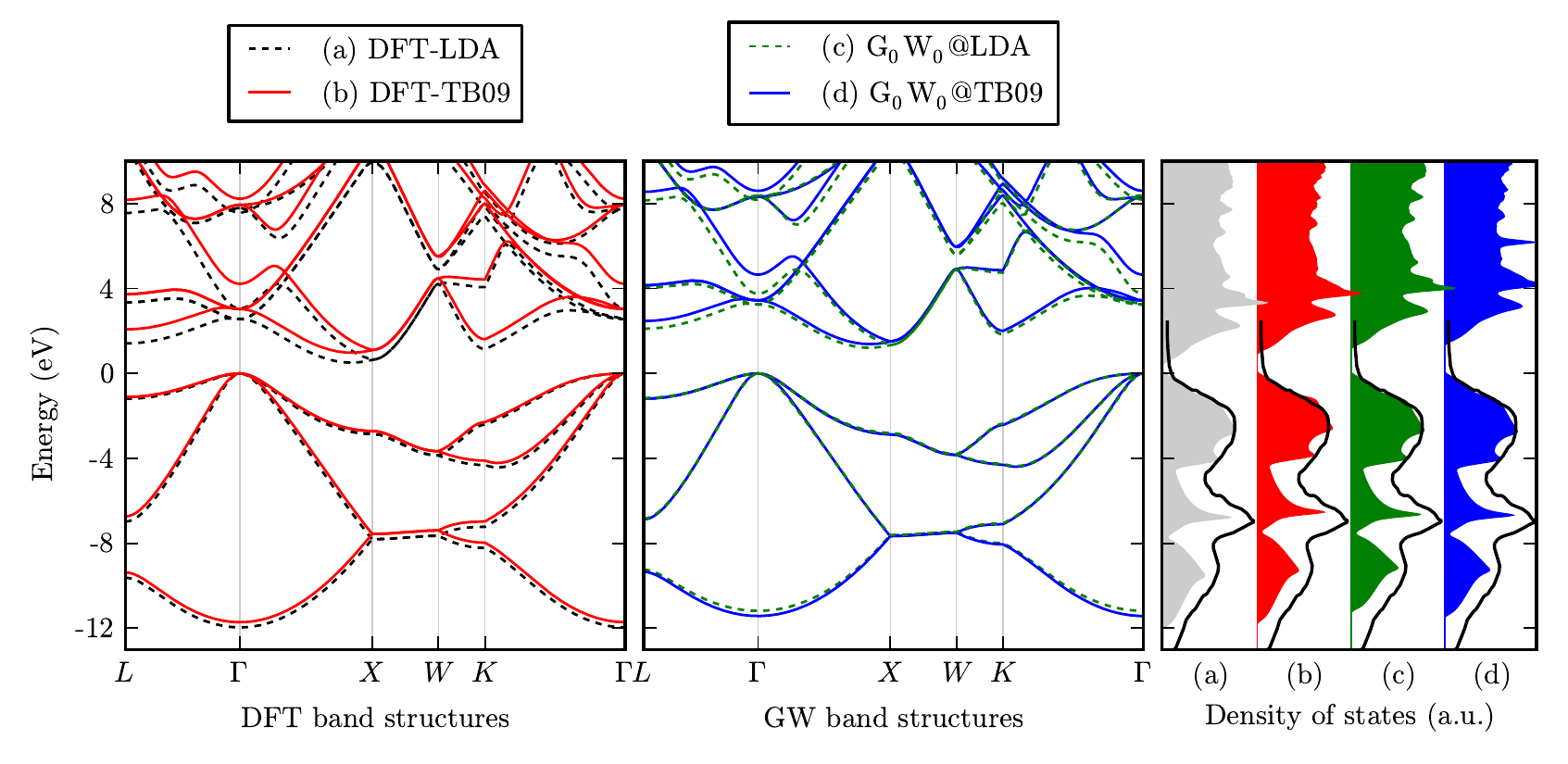}
\caption{\label{fig:silicon_bs}{(Color online) Band structure and DOS of bulk silicon in the diamond structure computed in DFT with (a) the LDA XC, (b) the TB09 XC and in one-shot $GW$ using (c) DFT-LDA and (d) DFT-TB09 as a starting point. In the left panel, the dashed (black) lines correspond to the DFT-LDA band structure and the full (red/gray) lines represent the DFT-TB09 band structure. In the middle panel, the dashed (green/light gray) lines correspond to the $G_0W_0$@LDA band structure and the full (blue/dark gray) lines represent the $G_0W_0$@TB09 band structure. In the right panel, the valence DOS (black line) is compared to XPS spectrum~\cite{Ley1972} in the right panel.}}
\end{figure*}

\begin{figure*}[ht!]
\center
\includegraphics{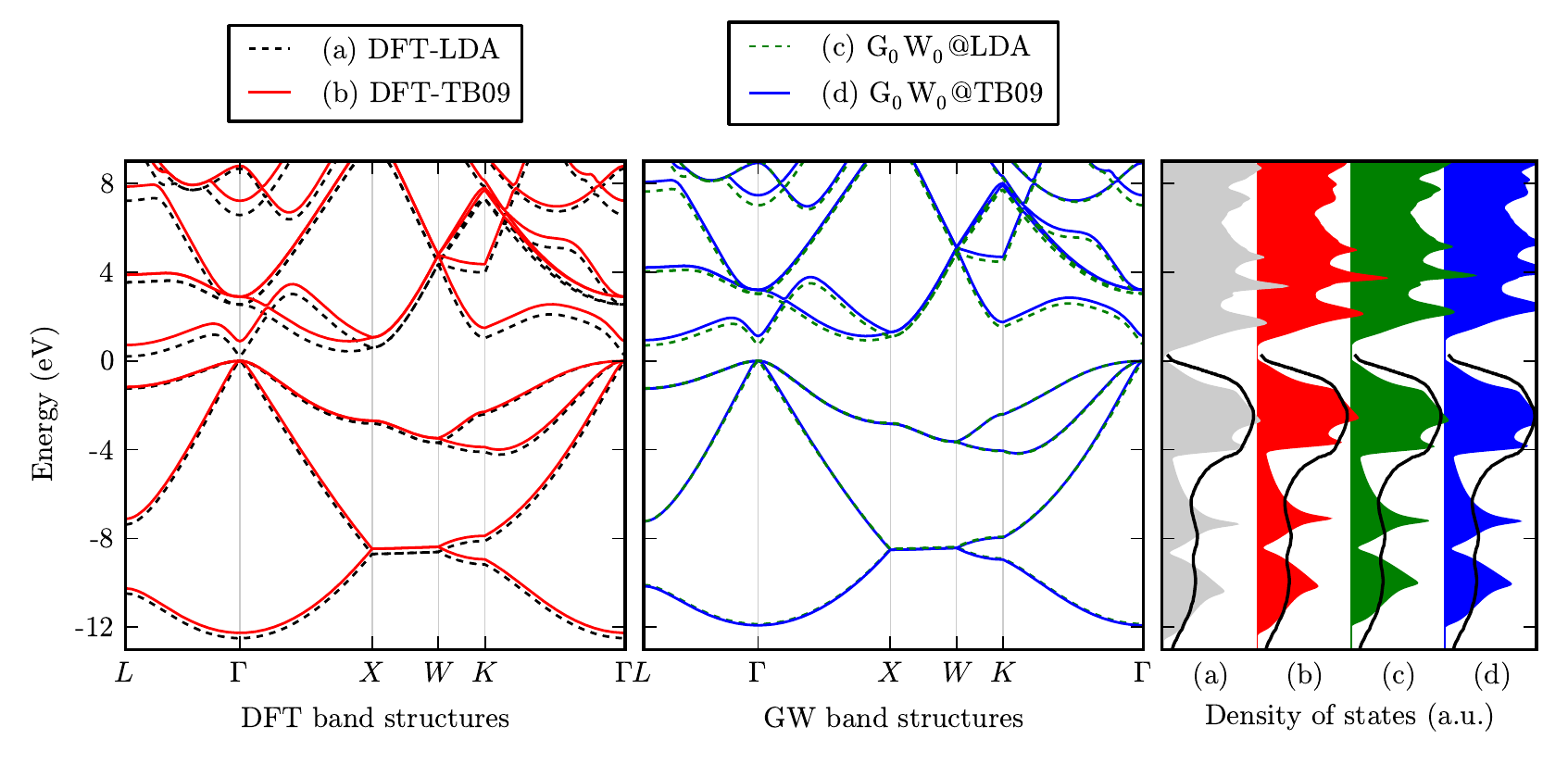}
\caption{\label{fig:germanium_bs}{(Color online) Band structure and DOS of bulk germanium in the diamond structure computed in DFT with (a) the LDA XC, (b) the TB09 XC and in one-shot $GW$ using (c) DFT-LDA and (d) DFT-TB09 as a starting point. The XPS spectrum is from Ref.~\onlinecite{Eastman1974}. The color scheme is the same as in Fig.~\ref{fig:silicon_bs}.}}
\end{figure*}

The silicon and germanium band structures and densities of states (DOS) are presented in Figs.~\ref{fig:silicon_bs} and \ref{fig:germanium_bs} respectively. The DOS are compared to experimental X-ray photoemission spectra~\cite{Ley1972,Eastman1974} (XPS). It is clear from the figures that both LDA and TB09 provide a rather fair description of the valence DOS. The main features of the band structure are indeed correctly reproduced by both XC approximations. The more elaborate $G_0W_0$-corrected DOS does not give rise to any major change in the valence electronic structure. Table~\ref{tab:bandgaps_all} gathers the direct and indirect band gap values while  Table~\ref{tab:bandwidths_all} gathers the valence band widths as well as first conduction band widths, obtained with the different methods. As expected, the band gaps obtained with TB09 are fundamentally improved over LDA. $G_0W_0$@TB09 opens the gap further, overshooting the experimental values. When compared to XPS results, our DFT results show that both XC functionals yield too small band widths. Further $G_0W_0$ correction even worsens this narrowing although the relative error with respect to experiment is reasonable in all four cases.

\begin{figure*}
\center
\includegraphics{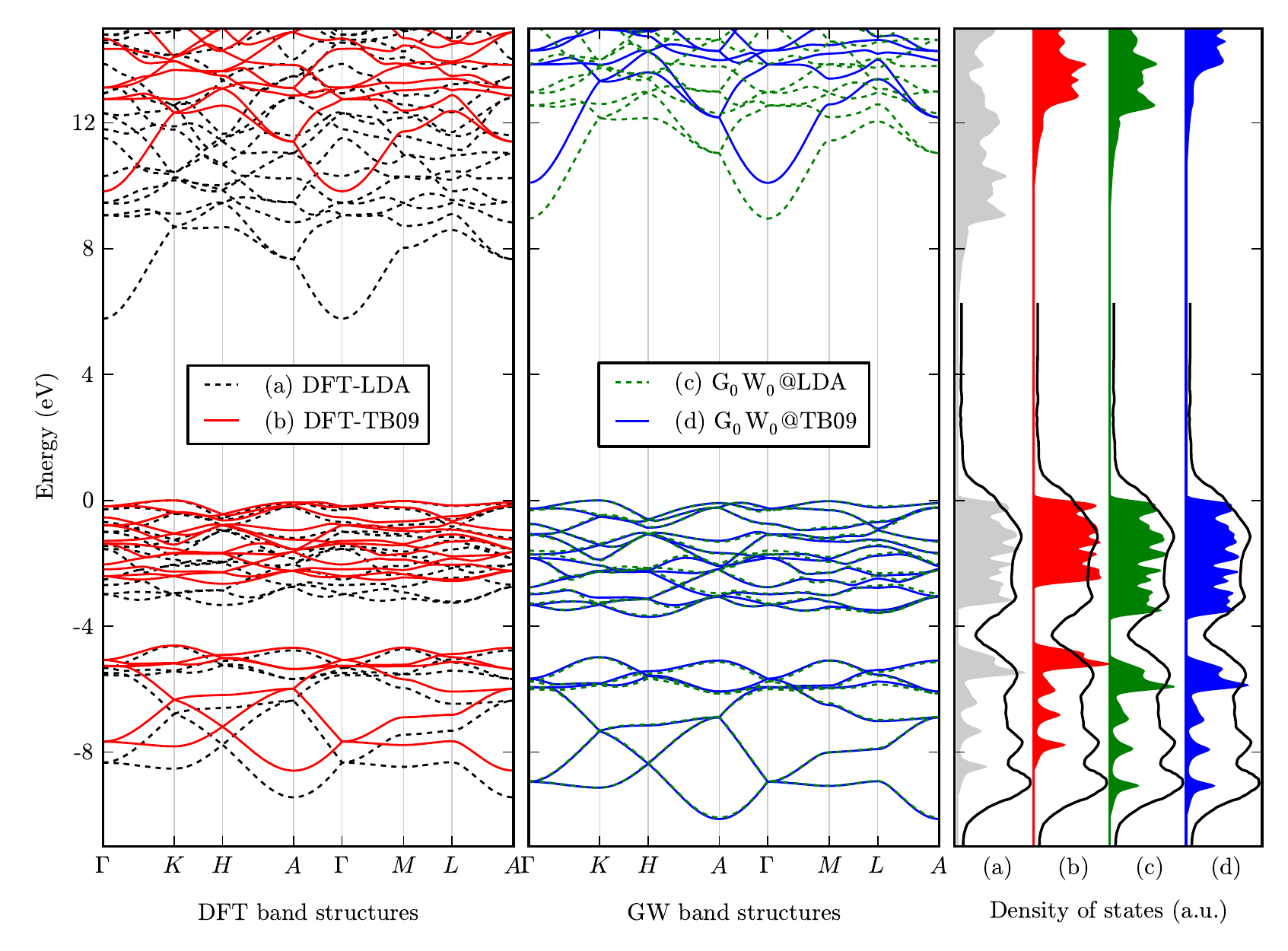}
\caption{\label{fig:quartz_bs}{(Color online) Band structure and density of states of $\alpha$-quartz computed in DFT with (a) the LDA XC, (b) the TB09 XC and in one-shot $GW$ using (c) DFT-LDA and (d) DFT-TB09 as a starting point. The XPS spectrum is from Ref.~\onlinecite{Zakaznova-Herzog2005}. The color scheme is the same as in Fig.~\ref{fig:silicon_bs}.}}
\end{figure*}

\subsection{Silicon dioxide}

Silicon dioxide is an insulator with a large experimental band gap of $\sim 9$ eV~\cite{Weinberg1979, Laughlin1980} where DFT in the LDA approximation underestimates the gap by $\sim 3$ eV. It is thus a more challenging case than Si and Ge, to test the accuracy of the electronic structure obtained with the Tran-Blaha functional. Here, the $\alpha$-quartz polymorph ($P3_221$) has been considered with the experimental lattice parameters and internal coordinates from Wyckoff~\cite{Wyckoff1963} ($a$=$4.91$~\AA, $c$=$5.40$~\AA, $u_{\mathrm{Si}}$=$0.465$, $x_{\mathrm{O}}$=$0.415$, $y_{\mathrm{O}}$=$0.272$, $z_{\mathrm{O}}$=$0.12$). An energy cut-off of $40$ Ha for plane waves and a 4$\times$4$\times$4 \textit{k}-point mesh have been used. Finally, the dielectric matrix was expanded with an energy cutoff of $8$ Ha and 600 bands were used for the computation of the $GW$ corrections.

Fig.~\ref{fig:quartz_bs} shows the band structures and DOS obtained with LDA, TB09, $G_0W_0$@LDA and $G_0W_0$@TB09. The fundamental gap is indirect, from $K$ to $\Gamma$. The valence bands are mainly composed of three groups. The lowest bands (not shown in the figure) lie around $20$ eV below the Fermi level. These bands are mainly composed of O $2s$-like orbitals. The second lowest group of bands, corresponding to bonding O $2p$-like orbitals and Si $3s$,$3p$-like orbitals, is located between $5$ and $10$ eV below the Fermi energy. Finally, the upper valence band group is composed of non-bonding O $2p$-like orbitals. The band gaps and different band widths of $\alpha$-quartz obtained with the different XC approximations, in DFT and $G_0W_0$ are reported in Table~\ref{tab:bandgaps_all} and Table~\ref{tab:bandwidths_all}. 

As expected, the TB09 fundamental and direct gaps are closer to the experimental values than the LDA ones and are similar to recent theoretical gaps~\cite{Kresse2012} using TB09 in the PAW method. Comparing the DOS with a recent XPS experiment,~\cite{Zakaznova-Herzog2005} it is clear that the TB09 drastically underestimates the valence band widths. Applying $G_0W_0$ corrections rectifies this error, leading to a $G_0W_0$@TB09 valence band structure very close to the $G_0W_0$@LDA one.

\begin{figure*}
\center
\includegraphics{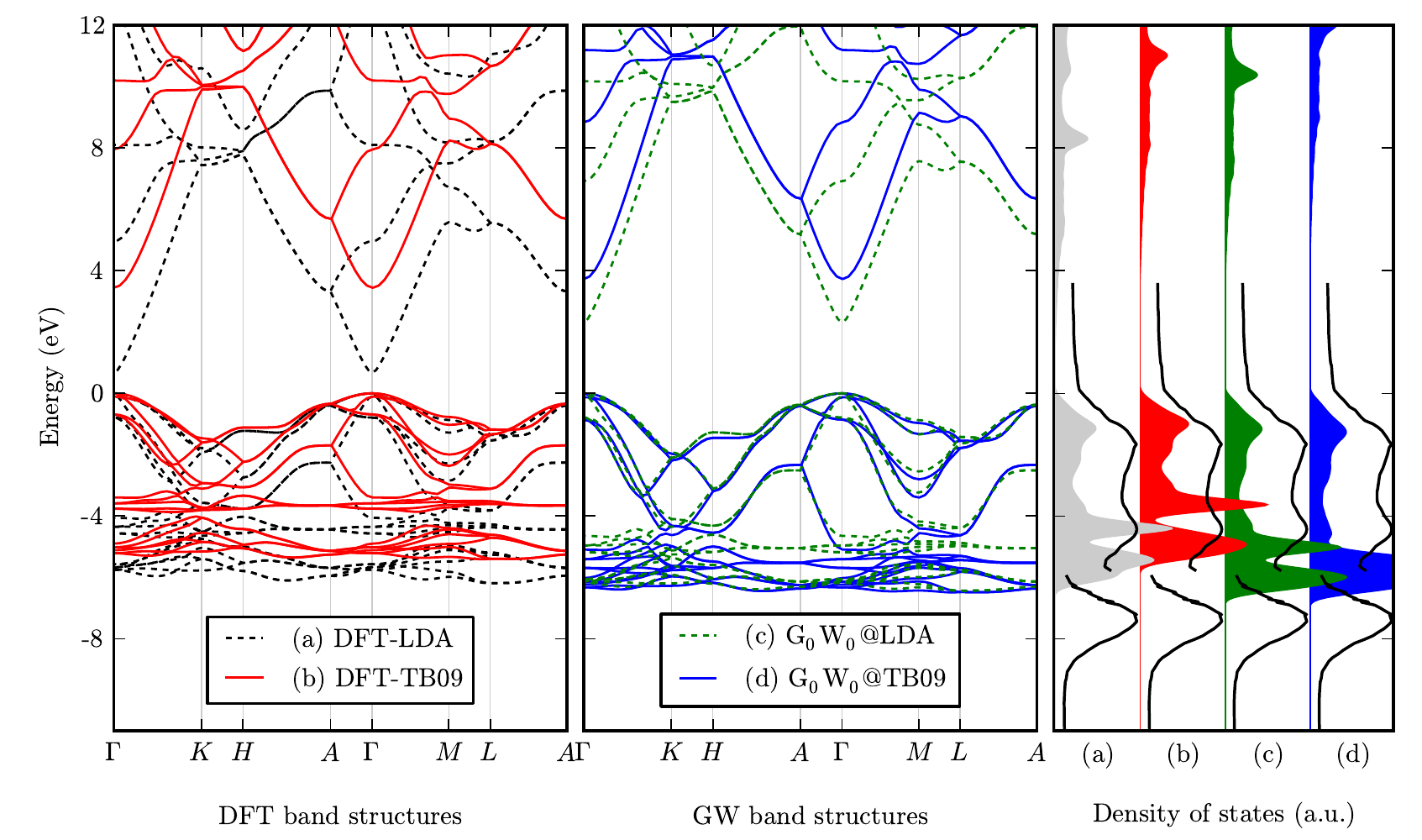}
\caption{\label{fig:zinc_oxide_bs}{(Color online) Band structure and density of states of ZnO computed in DFT with (a) the LDA XC, (b) the TB09 XC and in one-shot $GW$ using (c) DFT-LDA and (d) DFT-TB09 as a starting point. The XPS spectrum is from Ref.~\onlinecite{King2009}. The range of energies from -6 eV to the Fermi level has been magnified with respect to the rest of the XPS spectrum. The color scheme is the same as in Fig.~\ref{fig:silicon_bs}.}}
\end{figure*}

\subsection{Zinc oxide}

Zinc oxide (ZnO) is a widely studied material due to its application as a transparent conducting oxide. The wurtzite structure ($P6_3mc$ - $B4$) of ZnO has been considered here with the experimental lattice parameters from Kihara and Donnay.~\cite{Kihara1985} It is also a benchmark case for band gap predictions as there is no consensus on the theoretical gap obtained from DFT or $GW$ methods, with reported $GW$ values ranging from $2.1$ eV to $4.2$ eV (see Ref.~\onlinecite{Stankovski2011} and references therein).

The pseudopotential for the zinc atom includes the whole $n$=$3$ shell in the valence configuration. A $\Gamma$-centered 8$\times$8$\times$5 \textit{k}-point mesh has been used and the plane-wave energy cut-off used for the wavefunctions and dielectric matrix are $150$ Ha and $20$ Ha respectively. For the computation of the $GW$ corrections, 500 bands have been included with a compensation energy of 8 Ha for higher states. This is equivalent to the use of at least 3000 bands without the application of the extrapolar method.~\cite{Stankovski2011}

The band structures and DOS are given in Fig.~\ref{fig:zinc_oxide_bs}. The DOS is compared to XPS.~\cite{King2009} As already shown in other previous studies,~\cite{Lany2005,Laskowski2006,Lany2007,Zhou2008} LDA fails to provide a correct description of the Zn $3d$ levels and gives rise to a strong hybridization of these levels with the O and Zn $p$-states. The same authors suggest to add a Hubbard U term in order to lower the position of the {Zn} $3d$-states and actually decouple them from the $p$-states. This results at the same time in an increased band gap closer to measurements. In any case, the same hybridization problem is observed when using the TB09 even if a tiny internal band gap around -$4$ eV is obtained. Nevertheless, considering all these states as a single group, the valence band width is once more clearly narrowed with the TB09 ($5.41$ eV) compared to LDA ($6.19$ eV). The $G_0W_0$ corrected band structures are much closer to each others, at least for the highest $p$ states (from -4 eV to the Fermi level). In Table~\ref{tab:bandwidths_all}, the present theoretical results are compared to the experimental band width ($9$ eV)~\cite{Vogel1995,Ozgur2005} corresponding to the sum of $p$ states ($5.3$ eV), $d$ states ($2.5$ eV) and the separating internal band gap ($1.2$ eV). This experimental band width is also in good agreement with the XPS spectrum reported in Fig.~\ref{fig:zinc_oxide_bs}.

The TB09 yields a band gap value of $3.44$ eV which is again in much better agreement with experimental one ($3.6$ eV~\cite{Tsoi2006}) than the LDA ($0.67$ eV). The $G_0W_0$ correction to LDA leads to a larger band gap ($2.34$ eV), still too low with respect to the experimental value while starting from the TB09 electronic structure leads to a $G_0W_0$ gap of $3.73$ eV.

\begin{figure*}
\center
\includegraphics{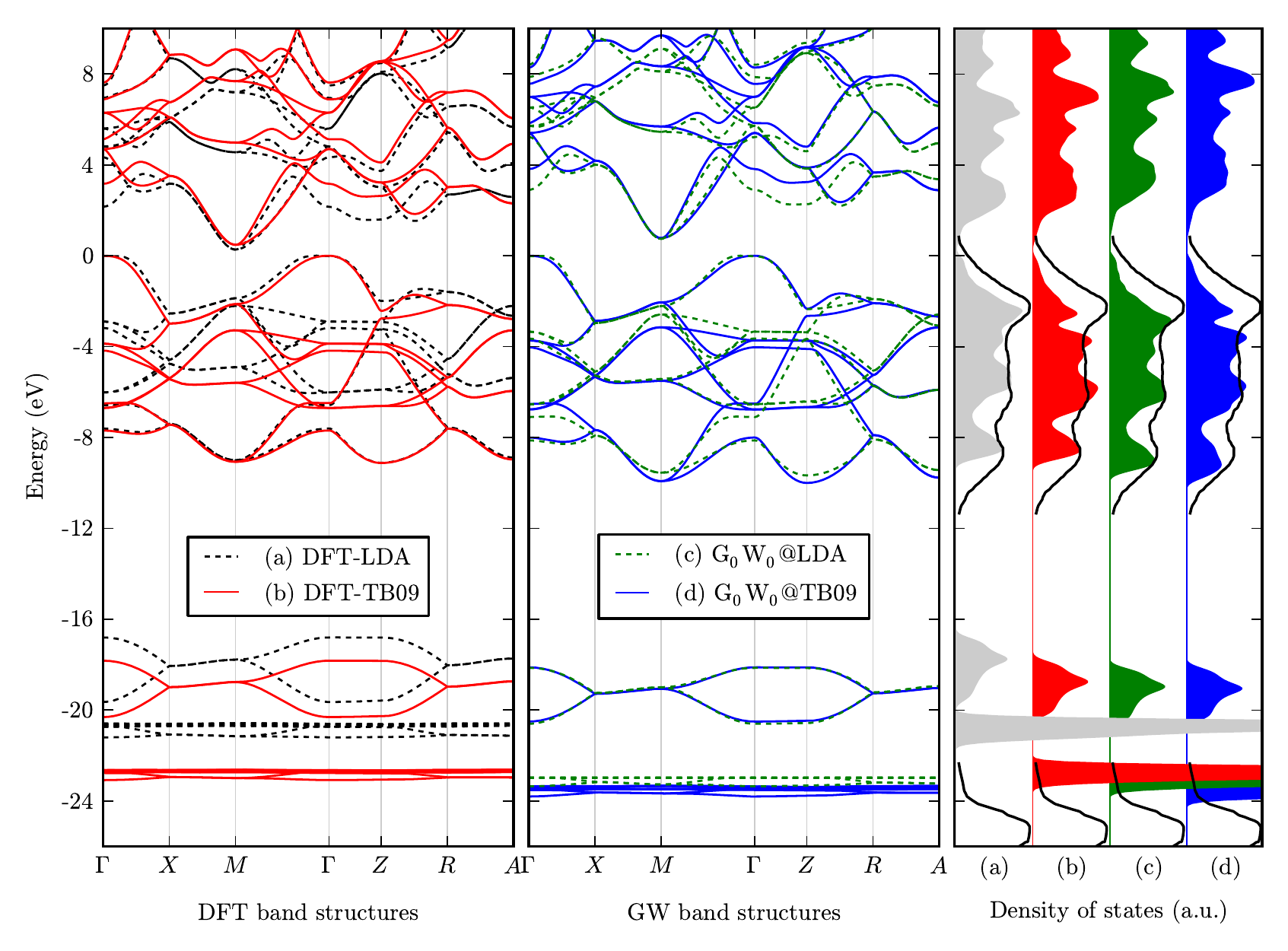}
\caption{\label{fig:tin_oxide_bs}{(Color online) Band structure and density of states of SnO computed in DFT with (a) the LDA XC, (b) the TB09 XC and in one-shot $GW$ using (c) DFT-LDA and (d) DFT-TB09 as a starting point. The XPS spectrum is from Ref.~\onlinecite{Themlin1992}. The color scheme is the same as in Fig.~\ref{fig:silicon_bs}.}}
\end{figure*}

\subsection{Tin oxides}

Two forms of tin oxides have been considered : stannous SnO and stannic SnO$_2$ oxide. Stannous oxide crystallizes in the $P4/nmm$ (B10) structure while stannic oxide adopts the rutile form ($P4_2/mnm$). Their experimental lattice parameters and internal coordinates are taken from Refs.~\onlinecite{Pannetier1980} and~\onlinecite{Haines1997} respectively. For both systems, an energy cutoff of $100$ Ha was used for the wavefunctions. The \textit{k}-point meshes used were 4$\times$4$\times$3 and 4$\times$4$\times$6 for SnO and SnO$_2$ respectively. For the $GW$ corrections, the dielectric matrix was expanded using a cutoff energy of $10$ Ha and 1000 bands have been used for the computation of the $GW$ corrections.

\begin{figure*}
\center
\includegraphics{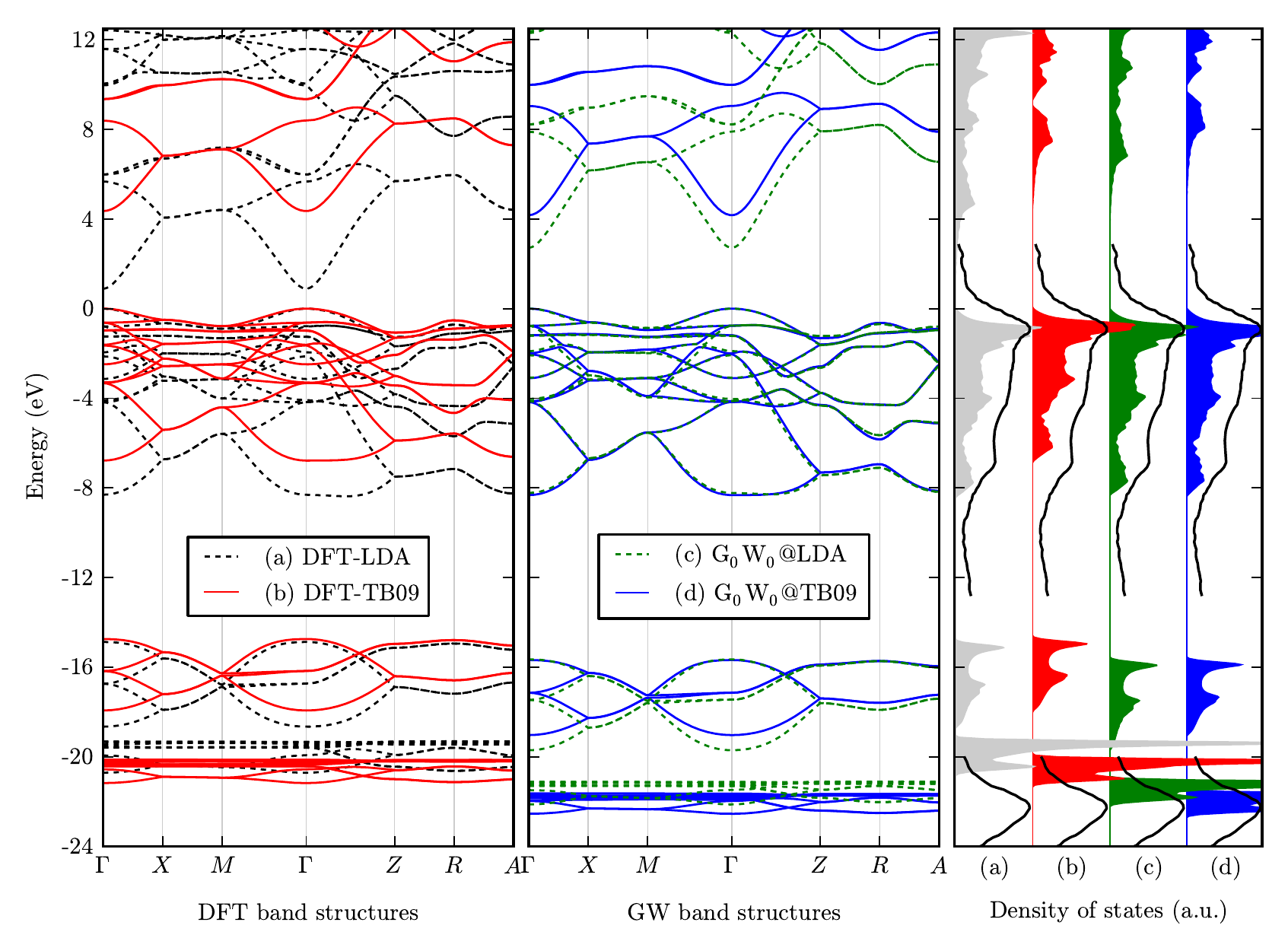}
\caption{\label{fig:tin_dioxide_bs}{(Color online) Band structure and density of states of SnO$_2$ computed in DFT with (a) the LDA XC, (b) the TB09 XC and in one-shot $GW$ using (c) DFT-LDA and (d) DFT-TB09 as a starting point. The XPS spectrum is from Ref.~\onlinecite{Themlin1992}. The color scheme is the same as in Fig.~\ref{fig:silicon_bs}.}}
\end{figure*}

The band structures and DOS are given in Figs.~\ref{fig:tin_oxide_bs} and \ref{fig:tin_dioxide_bs}. The highest group of valence bands of both oxides is rather well described even if some discrepancies appear between LDA and TB09, especially for the band width in case of SnO$_2$. For SnO, it is clear that the highly localized $d$ states (lying at $\sim$23-24 eV below the Fermi level) are better described with the TB09 as they are pushed down in energy, closer to the $G_0W_0$ corrected ones. The O $s$-states (around $\sim$18-20 eV below the Fermi level) are better positioned with TB09 than with LDA as confirmed by the $G_0W_0$ results. In contrast with SiO$_2$ and ZnO, but similarly to Si and Ge, the valence band widths in SnO are not narrowed with the TB09 leading to a surprisingly good description of the valence states.

The fundamental band gap of SnO is indirect from $\Gamma$ to $M$. TB09 leads to larger fundamental and direct band gaps than LDA but both DFT fundamental band gaps are still lower than the measured values. The addition of a $G_0W_0$ leads to indirect band gaps in very close agreement with the experiment whereas the direct gap at $\Gamma$ is overestimated by $G_0W_0$@TB09.

For tin dioxide, the flat bands corresponding to $d$ states (around -22 eV to -20 eV) are also somewhat better positioned in energy with the TB09 than within LDA. The group of bands around -16 eV to -18 eV which is composed of O $s$ states and some Sn $s$ and $p$ states is clearly shrunk with the TB09 compared to the LDA and $G_0W_0$ structures. The highest valence bands are also strongly contracted with TB09.

Concerning the band gap, it is clearly underestimated within DFT-LDA while the TB09 leads to a value larger than the experimental one.~\cite{Batzill2005} The $G_0W_0$@LDA band gap is still lower than in experiments while the $G_0W_0$@TB09 theoretical gap is closer to the measured value.

\subsection{Calcium sulfide}

Calcium sulfide has been studied in the rock salt structure ($Fm\overline{3}m$ - $B1$) using the experimental lattice parameters.~\cite{Wyckoff1960} The energy cut-off used for the wavefunctions was $45$ Ha and the reciprocal space was discretized using $29$ \textit{k}-points in the irreducible Brillouin zone. The dielectric matrix was expanded using an energy cut-off of $10$ Ha and 200 bands have been used for the calculation of the $GW$ corrections.

The band structures and DOS are given in Fig.~\ref{fig:calcium_sulfide_bs}. The fundamental band gap is indirect from $\Gamma$ to $X$. The TB09 leads to increased values of the band gaps as shown in Table~\ref{tab:bandgaps_all} while the LDA underestimates the experiments. The $G_0W_0$@LDA band gaps are closer to experiment. $G_0W_0$@TB09 strongly overestimates the direct band gap at $\Gamma$ while the direct band gap at $X$ and the indirect band gap are within $5\%$ of the experimental values. Unfortunately, no QS$GW$ results were available in literature to compare with for this large band gap semiconductor.

The upper valence band composed of sulfur $p$-states is again clearly narrowed by TB09. In contrast, LDA is in much better agreement with the $G_0W_0$ values. The band widths obtained with $G_0W_0$@LDA and $G_0W_0$@TB09 are in close agreement with each other.

\subsection{Calcium oxide}

Calcium oxide CaO has been investigated in the rock salt structure ($Fm\overline{3}m$ - $B1$). The experimental lattice parameters~\cite{Smith1968} have been used. An energy cut-off of $38$ Ha and $29$ \textit{k}-points in the irreducible Brillouin zone have been used for the wavefunctions. The dielectric matrix was expanded using an energy cut-off of $10$ Ha and 380 bands have been used for the calculation of the $GW$ corrections.

\begin{figure*}[htp!]
\center
\includegraphics[width=\textwidth]{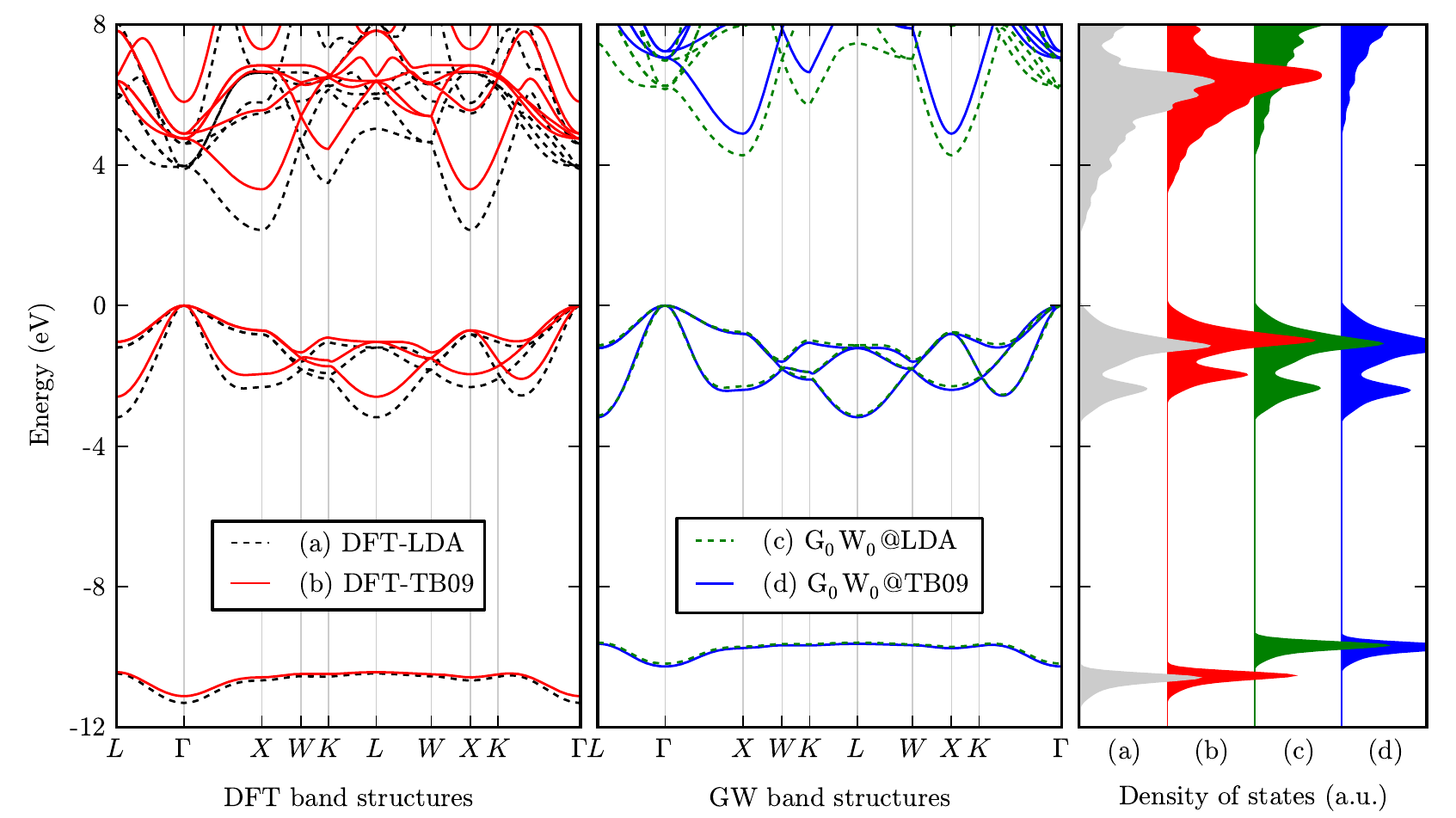}
\caption{\label{fig:calcium_sulfide_bs}{(Color online) Band structure and density of states of CaS computed in DFT with (a) the LDA XC, (b) the TB09 XC and in one-shot $GW$ using (c) DFT-LDA and (d) DFT-TB09 as a starting point. The color scheme is the same as in Fig.~\ref{fig:silicon_bs}.}}
%
\vspace{0.6cm}
\center
\includegraphics[width=\textwidth]{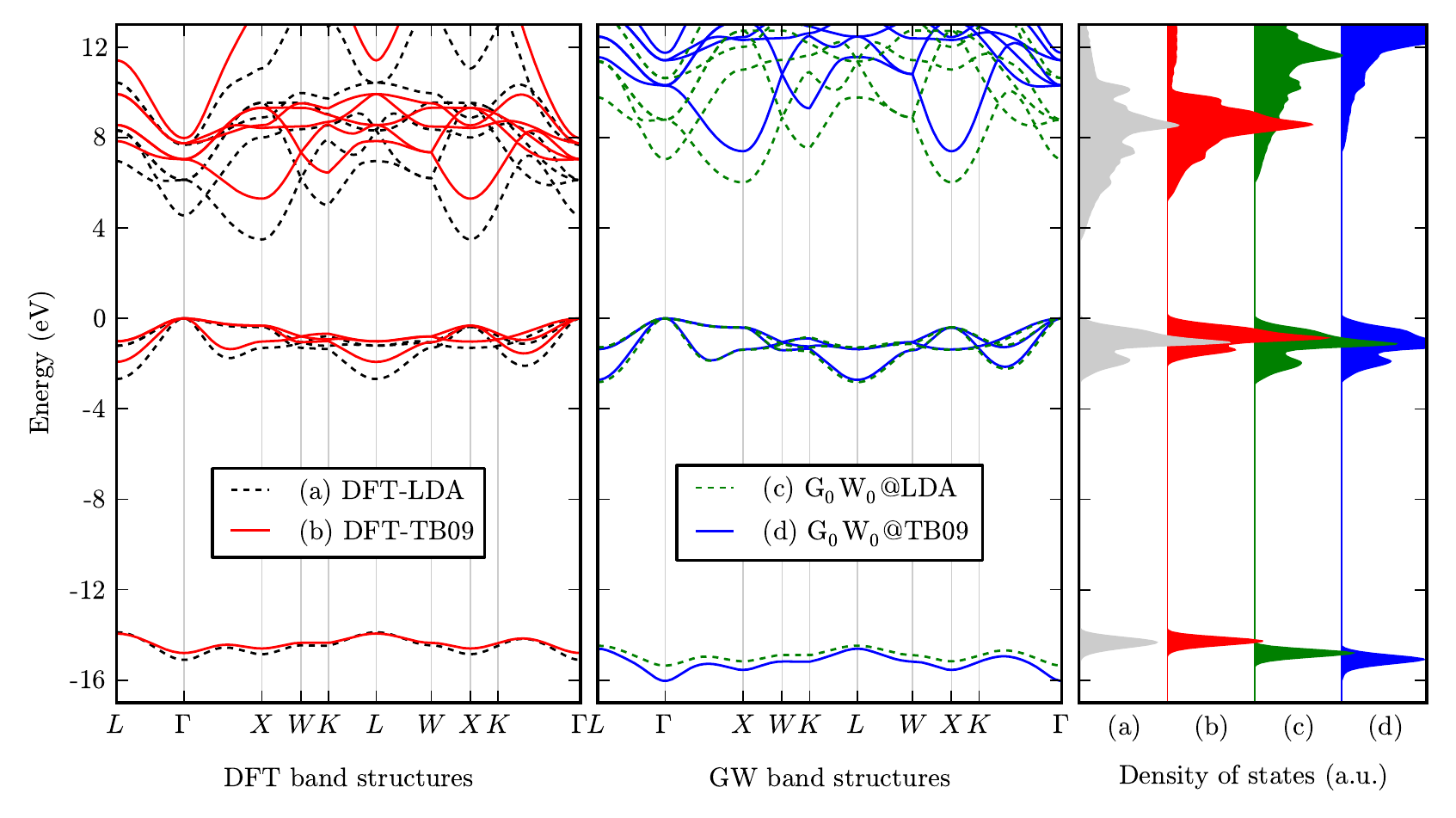}
\caption{\label{fig:calcium_oxide_bs}{(Color online) Band structure and density of states of CaO computed in DFT with (a) the LDA XC, (b) the TB09 XC and in one-shot $GW$ using (c) DFT-LDA and (d) DFT-TB09 as a starting point. The color scheme is the same as in Fig.~\ref{fig:silicon_bs}.}}
\end{figure*}

\begin{figure*}
\center
\hspace*{-0.1cm}
\includegraphics{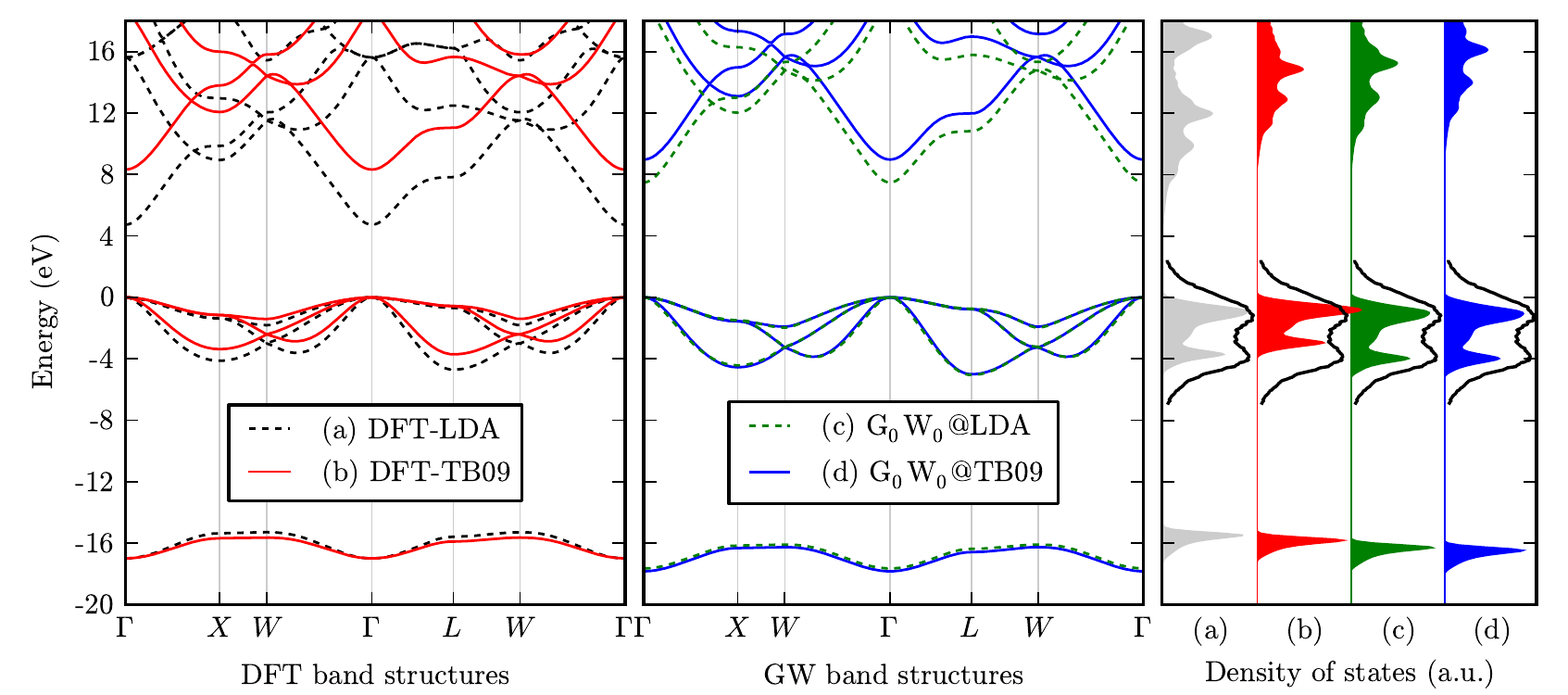}
\caption{\label{fig:magnesium_oxide_bs}{(Color online) Band structure and density of states of MgO computed in DFT with (a) the LDA XC, (b) the TB09 XC and in one-shot $GW$ using (c) DFT-LDA and (d) DFT-TB09 as a starting point. The XPS spectrum is from Ref.~\onlinecite{French1990}. The color scheme is the same as in Fig.~\ref{fig:silicon_bs}.}}
\end{figure*}

The band structures and DOS obtained are shown in Fig.~\ref{fig:calcium_oxide_bs}. The obtained conduction band structure is somewhat different with the LDA and the TB09. In LDA, there is a local minimum in the conduction band at $\Gamma$ while this band is pushed upwards with the TB09. In both cases, the gap is indirect from $\Gamma$ to $X$. The band gaps obtained are gathered in Table \ref{tab:bandgaps_all}. There is some controversy on the exact value and on the indirect character of the fundamental gap such that no reliable experimental value could be compared with our results.

Concerning the valence electronic structure, the upper valence bands are once more narrowed by TB09 when compared to LDA or $G_0W_0$-corrected band structures.

\subsection{Magnesium oxide}

The most stable phase of magnesium oxide (MgO) is in the rock-salt structure ($Fm\overline{3}m$). The experimental~\cite{Bobade2012} lattice parameter $a$=$4.203$~\AA has been used. The fundamental gap, which is direct at $\Gamma$, amounts to $7.83$ eV.~\cite{Whited1973} The wavefunctions were expanded using $80$ Ha as the energy cutoff for the plane waves and $44$ \textit{k}-points in the irreducible Brillouin zone. An energy cut-off of $16$ Ha has been used for the plane wave representation of the dielectric matrix while 300 bands have been included for the computation of the $GW$ corrections.

The band structures and DOS are shown in Fig.~\ref{fig:magnesium_oxide_bs}. The band gap is underestimated within DFT-LDA while the TB09 leads to a value larger than the experimental value. The $G_0W_0$@LDA band gap is closer to experiments while $G_0W_0$@TB09 pushes the gap even further away from the measured value. In fact, the $G_0W_0$@TB09 gap is very close to the reported QS$GW$ result.

The valence bands are clearly shrunk when using the TB09. Comparing the DOS with XPS experiments~\cite{French1990} shows evidently that the distance between the two peaks in the DOS of the upper valence band is much smaller than the experimental value when using the TB09 while the LDA and the $G_0W_0$-corrected band structures are in much better agreement. The valence band widths and band gap values obtained with the different methods are collected in Table~\ref{tab:bandgaps_all} and Table~\ref{tab:bandwidths_all}.

\subsection{Lithium fluoride}

Lithium fluoride (LiF) has been studied in the rock-salt structure ($Fm\overline{3}m$) using the experimental lattice parameter~\cite{Ott1926} $a$=$4.028$~\AA. The wavefunctions were expanded using $40$ Ha as the energy cutoff for the plane waves and $29$ \textit{k}-points in the irreducible Brillouin zone. A kinetic energy cut-off of $16$ Ha has been used for the plane wave representation of the dielectric matrix and 500 bands have been used for the computation of the $GW$ corrections.

\begin{figure*}
\center
\hspace*{-0.1cm}
\includegraphics{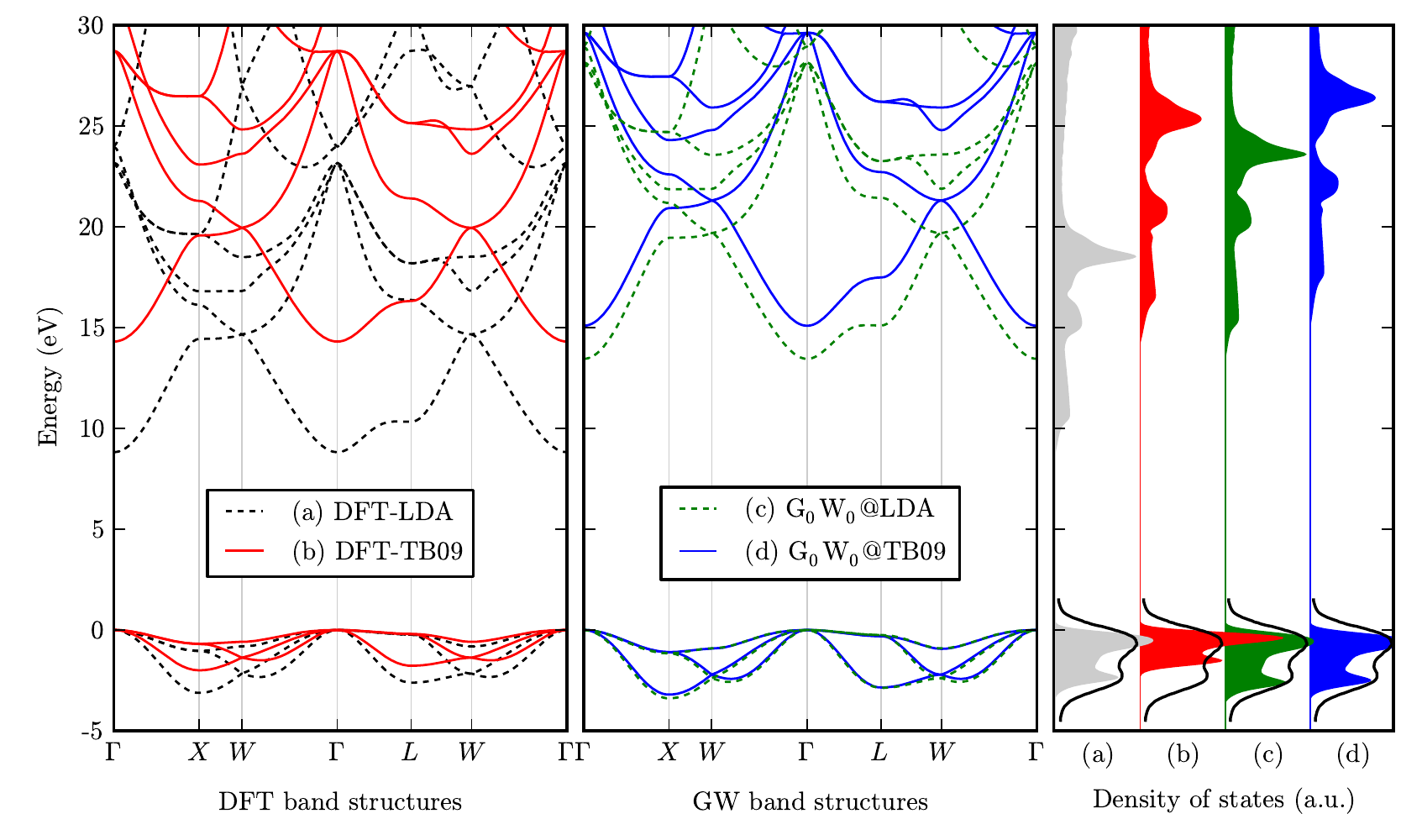}
\caption{\label{fig:lithium_fluoride_bs}{(Color online) Band structure and density of states of LiF computed in DFT with (a) the LDA XC, (b) the TB09 XC and in one-shot $GW$ using (c) DFT-LDA and (d) DFT-TB09 as a starting point. The XPS spectrum is from Ref.~\onlinecite{Himpsel1992}. The color scheme is the same as in Fig.~\ref{fig:silicon_bs}.}}
\end{figure*}

The band structures and DOS are shown in Fig.~\ref{fig:lithium_fluoride_bs}. There is a strong narrowing of the valence band when using the TB09. Comparing the DOS with XPS experiments~\cite{Himpsel1992} shows obviously that the valence band is much smaller than the experimental value when using the TB09 while the LDA and the $G_0W_0$-corrected band structures are in better agreement. The two-peak structure in the XPS spectrum is also much better rendered with the LDA DOS whereas the TB09 is far from the experiment. Both $G_0W_0$-DOS fairly reproduce this peak structure.

The valence band width and band gap values obtained with the different methods are gathered in Table~\ref{tab:bandgaps_all} and Table~\ref{tab:bandwidths_all}. As for the other materials, the band gap is underestimated within DFT-LDA while the TB09 corrects this, yielding a value of $14.31$ eV in much closer agreement with the experimental value~\cite{Piacentini1976} of $14.2$ eV. The $G_0W_0$ correction opens the LDA band gap closer to experiment while $G_0W_0$@TB09 opens the gap even further, overshooting the experimental value by almost $1$eV. It is again very close to the QS$GW$ band gap.

\begin{table}
\caption{\label{tab:bandgaps_all} Fundamental (E$_\mathrm{g}$) and direct (E$_\mathrm{g,d}$) band gaps of all the materials in this study.}
\renewcommand{\arraystretch}{1.1}
\begin{ruledtabular}
\begin{tabular}{llrrrrrd}
& & \multicolumn{2}{c}{DFT} & \multicolumn{2}{c}{$G_0W_0$} & & \\
& & \multicolumn{1}{c}{LDA} & \multicolumn{1}{c}{TB09} & \multicolumn{1}{c}{@LDA} & \multicolumn{1}{c}{@TB09} & \multicolumn{1}{c}{QS$GW$} & \multicolumn{1}{c}{Expt.} \\
\hline
Si & E$_\mathrm{g}$ & 0.51 & 0.98 & 1.21 & 1.38 & 1.24\footnotemark[1]\footnotetext[1]{Ref.~\onlinecite{Kresse2012}} & 1.12\footnotemark[2]\footnotetext[2]{Ref.~\onlinecite{Kasper2005}} \\
\vspace{1mm} & E$_\mathrm{g,d}$ & 2.56 & 3.04 & 3.25 & 3.44 & 3.30\footnotemark[1] & 3.20\footnotemark[2] \\
Ge & E$_\mathrm{g}$ & 0.20 & 0.71 & 0.70 & 0.93 & 0.95\footnotemark[3]\footnotetext[3]{Ref.~\onlinecite{Shishkin2007a}} & 0.66\footnotemark[2] \\
\vspace{1mm} & E$_\mathrm{g,d}$ & 0.23 & 0.89 & 0.73 & 1.12 &  & 0.80\footnotemark[2] \\
SiO$_2$ & E$_\mathrm{g}$ & 5.77 & 9.82 & 8.96 & 10.09 & 9.7\footnotemark[1]\1 & 8.9\footnotemark[4]\footnotetext[4]{Ref.~\onlinecite{Laughlin1980}} \\
\vspace{1mm} & E$_\mathrm{g,d}$ & 6.06 & 10.01 & 9.27 & 10.36 & 10.1\footnotemark[1]\1 & 10.0\footnotemark[5]\footnotetext[5]{Ref.~\onlinecite{Chang2000}} \\
ZnO\vspace{1mm} & E$_\mathrm{g}$ & 0.67 & 3.44 & 2.32 & 3.73 & 3.8\footnotemark[3]\1 & 3.6\footnotemark[6]\footnotetext[6]{Ref.~\onlinecite{Tsoi2006}} \\
SnO & E$_\mathrm{g}$ & 0.27 & 0.48 & 0.74 & 0.78 & 1.38\footnotemark[7]\footnotetext[7]{this work, obtained using a $G_0W_0$@scCOHSEX approach (for a description of the method, see Ref. \onlinecite{Bruneval2006})} & 0.7\footnotemark[8]\footnotetext[8]{Ref.~\onlinecite{Ogo2008}} \\
\vspace{1mm} & E$_\mathrm{g,d}$ & 2.17 & 3.17 & 2.92 & 3.82 & 3.88\footnotemark[7] & 2.77\footnotemark[1]\\
SnO$_2$\vspace{1mm} & E$_\mathrm{g}$ & 0.89 & 4.35 & 2.72 & 4.17 & 4.28\footnotemark[7] & 3.6\footnotemark[9]\footnotetext[9]{Ref.~\onlinecite{Batzill2005}} \\
CaS & E$_\mathrm{g}$ & 2.15 & 3.31 & 4.28 & 4.89 & & 4.43\footnotemark[10]\footnotetext[10]{Ref.~\onlinecite{Kaneko1990}} \\
    & E$_{\mathrm{g,d,}\Gamma}$ & 3.89 & 4.76 & 5.57 & 7.06 &  & 5.80\footnotemark[10] \\
\vspace{1mm} & E$_{\mathrm{g,d,}X}$ & 2.97 & 4.01 & 5.13 & 5.77 &  & 5.34\footnotemark[10]\\
CaO & E$_\mathrm{g}$ & 3.49 & 5.30 & 6.02 & 7.39 & 7.57\footnotemark[11]\footnotetext[11]{Ref.~\onlinecite{Schilfgaarde2006}} & 7.0\footnotemark[11] \\
    & E$_{\mathrm{g,d,}\Gamma}$ & 4.55 & 7.04 & 6.49 & 10.36 &  & 7.0\footnotemark[12]\footnotetext[12]{Ref.~\onlinecite{Whited1969,Whited1973}} \\
\vspace{1mm} & E$_{\mathrm{g,d,}X}$ & 3.87 & 5.62 & 6.46 & 7.85 &  & 7.3\footnotemark[13]\footnotetext[13]{Ref.~\onlinecite{Whited1973}}\\
MgO\vspace{1mm} & E$_\mathrm{g}$ & 4.73 & 8.32 & 7.48 & 8.97 & 9.16\footnotemark[3] & 7.83\footnotemark[13]\\
LiF & E$_\mathrm{g}$ & 8.82 & 14.31 & 13.45 & 15.09 & 15.9\footnotemark[3]\1 & 14.2\footnotemark[14]\footnotetext[14]{Ref.~\onlinecite{Piacentini1976}} \\
\end{tabular}
\end{ruledtabular}
\end{table}

\begin{figure}[ht!]
\center
\hspace*{-0.1cm}
\includegraphics[width=\columnwidth]{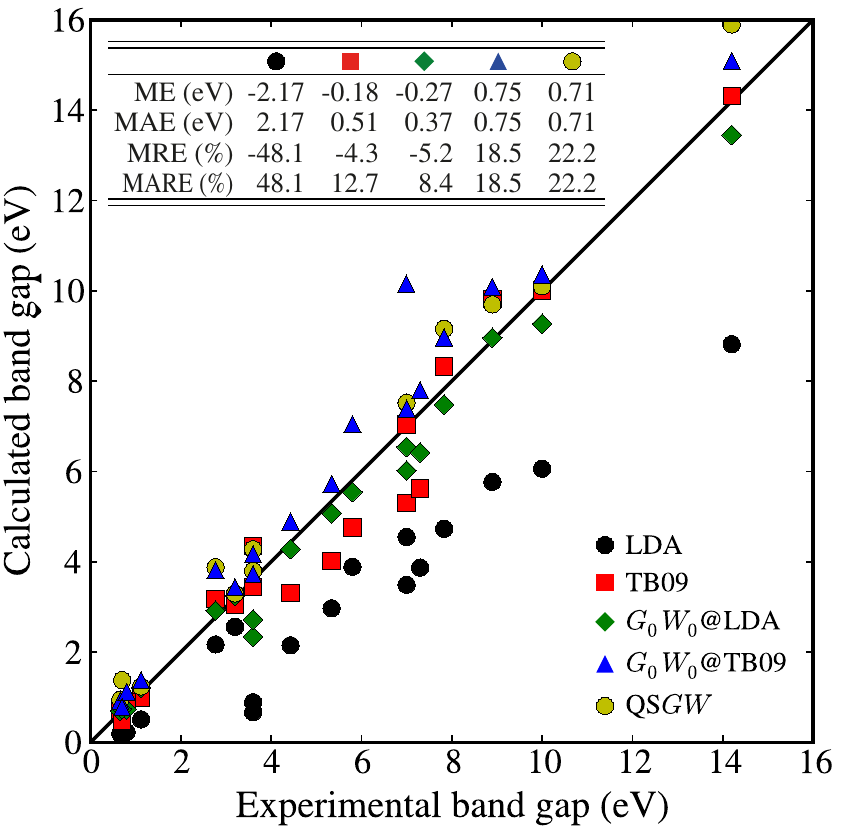}
\caption{\label{fig:bandgaps_all}{(Color online) Comparison of the theoretical and experimental band gaps (in eV) for all the different materials considered in this study. The different XC approximations for both DFT and $GW$ calculations are given as black dots (DFT-LDA), red squares (DFT-TB09), green diamonds ($G_0W_0$@LDA), blue triangles ($G_0W_0$@TB09), and yellow dots (QS$GW$). Corresponding band gap values are also reported in Table~\ref{tab:bandgaps_all}. Inset: Table of the mean error (ME), the mean absolute error (MAE), the mean relative error (MRE, in \%), and the mean absolute relative error (MARE, in \%) for the various XC functionals.} }
\end{figure}

\section{\label{sec:discussion}Discussion}

\subsection{Band gaps}

Fig.~\ref{fig:bandgaps_all} shows a comparison of all the theoretical band gaps calculated within this study with the experimental band gaps. It is clear that TB09 predicts band gaps in much better agreement with experiment than LDA. This is not surprising as this XC functional was especially designed for that purpose. Except for CaS and CaO, for which the values of the band gaps are not well reproduced, the error is on the order of $5$ to $10$\%. 

The band gaps obtained with $G_0W_0$@LDA are in better agreement with experiments than their DFT counterparts. On the other hand, $G_0W_0$@TB09 leads to an overestimation of the band gaps for all the compounds of the present study. Table~\ref{tab:bandgaps_all} also compares our $G_0W_0$ results to published values using Quasi-particle Self-consistent $GW$ (QS$GW$). Both QS$GW$ and $G_0W_0$@TB09 tend to overestimate the experimental band gaps. This is in line with the usual trend of QS$GW$ to overestimate band gaps. In this respect, $G_0W_0$@TB09 results thus seem closer to QS$GW$ ones. To confirm this, one would have to explicitly compare the quasi-particle and TB09 wavefunctions but this is beyond the scope of this work.

\subsection{Band widths}

\begin{figure}[ht!]
\center
\hspace*{-0.1cm}
\includegraphics[width=\columnwidth]{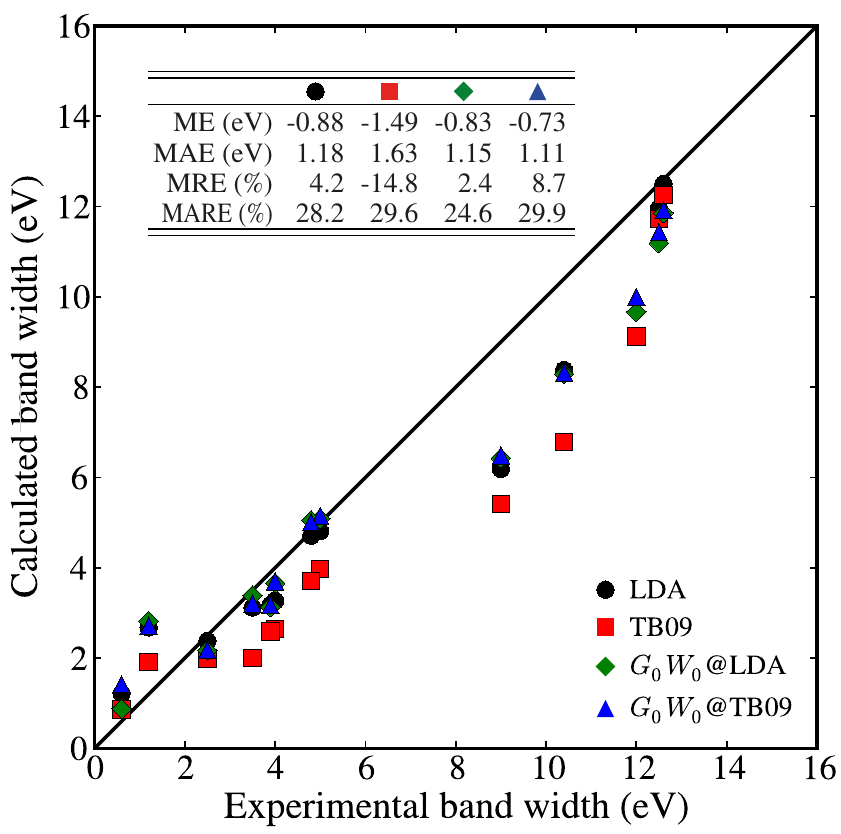}
\caption{\label{fig:bandwidths_all}{(Color online) Comparison of the theoretical and experimental band widths (in eV) for all the different materials considered in this study. The different XC approximations for both DFT and GW calculations are given as black dots (DFT-LDA), red squares (DFT-TB09), green diamonds ($G_0W_0$@LDA), and blue triangles ($G_0W_0$@TB09). Corresponding band width values are also reported in Table~\ref{tab:bandwidths_all}. Inset: Table of the mean error (ME), the mean absolute error (MAE), the mean relative error (MRE, in \%), and the mean absolute relative error (MARE, in \%) for the various XC functionals.} }
\vspace*{0.5cm}
\end{figure}

Conversely, the occupied band widths of most compounds are contracted when using the TB09 instead of the LDA (except for materials with small band gaps, namely Si, Ge and SnO). Comparisons with XPS spectra and experimental valence band widths clearly indicate that this characteristic of the band structure is not described correctly by the TB09. See for instance Fig.~\ref{fig:bandwidths_all} which compares all theoretical band widths to experimental band widths. As a matter of fact, the TB09 leads to a stronger localization than LDA, the latter thus being more suitable than TB09 to predict band widths at the DFT level. Indeed, band widths obtained with LDA are in closer agreement with both experimental and more elaborate $G_0W_0$ results. Finally, the valence electronic structures obtained with $G_0W_0$@LDA and $G_0W_0$@TB09 usually present similar electronic dispersions suggesting that the band narrowing obtained with TB09 tends to be corrected when using $G_0W_0$. 

\begin{figure}[h!]
\center
\hspace*{-0.1cm}
\includegraphics{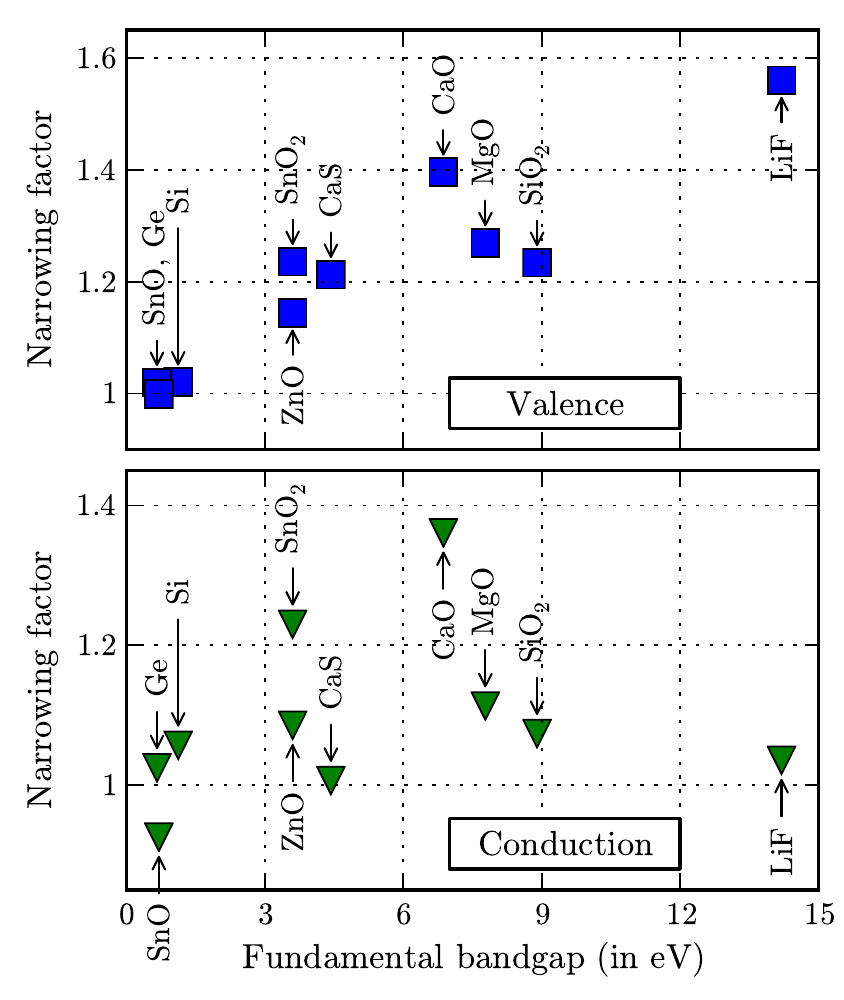}
\caption{\label{fig:narrowing_factor}{(Color online) Graph of the \textit{narrowing factor} of the upper valence band ($\mathrm{VBW}_{\mathrm{LDA}}/\mathrm{VBW}_{\mathrm{TB09}}$) and of the first conduction band ($\mathrm{1CBW}_{\mathrm{LDA}}/\mathrm{1CBW}_{\mathrm{TB09}}$) as a function of the fundamental band gap E$_\mathrm{g}$ of the materials considered in the present study. Materials with larger band gaps exhibit a stronger narrowing of the valence bands.}}
\end{figure}

Defining the \textit{narrowing factor} as the ratio of the upper valence band width in LDA to the upper valence band width in TB09, the experimental fundamental gap of the materials shows an overall linear correlation to the narrowing factor as shown in the upper panel of Fig.~\ref{fig:narrowing_factor}. It is indeed clear that the narrowing of these bands is more pronounced for materials with wider band gaps. This gives some hint on how the TB09 leads to larger gaps. Indeed, exact exchange methods such as Hartree-Fock lead to an overestimation of the band gaps. Similarly the TB09 opens the gaps by mixing in more exchange in the system thanks to the adjustment of the $c$ parameter. As a consequence of a more attractive exchange potential, the electrons are more localized and their electronic bands are narrowed.

Concerning the unoccupied bands, one possible measure of their dispersion behavior is the width of the first conduction band (1CBW). Such a quantity (the energy difference between the minimum and the maximum over the Brillouin zone of the lowest conduction band) might in principle be obtained from angle resolved inverse photoemission spectroscopy. However, in this work, we use this quantity to analyze theoretical results. Similarly to the valence bands, the values for the 1CBW reported in Table~\ref{tab:bandwidths_all} suggest a narrowing of the unoccupied bands within TB09 though not as systematic as for the valence bands, as shown in the lower panel of Fig.~\ref{fig:narrowing_factor}.

There has been a lot of interest in this XC functional and the TB09 has so far been applied to many systems. These studies have been conducted without checking that the ground state electronic structure results are coherent with more accurate MBPT approaches as done in this study. Because (a) the band widths are narrowed and (b) there is no energy functional from which the potential derives, this calls into question the reliability of other properties derived from the TB09-DFT ground-state. For instance, defect energy levels within the band gap or in the valence states might be misplaced.

\begin{table}
\caption{\label{tab:bandwidths_all} Valence band widths (VBW) and first conduction band widths (1CBW) of all the materials in this study. See the text for the definition of the 1CBW.}
\renewcommand{\arraystretch}{1.1}
\begin{ruledtabular}
\begin{tabular} {llddddd}
& & \multicolumn{2}{c}{DFT} & \multicolumn{2}{c}{$G_0W_0$} & \\
& & \multicolumn{1}{c}{LDA} & \multicolumn{1}{c}{TB09} & \multicolumn{1}{c}{@LDA} & \multicolumn{1}{c}{@TB09} & \multicolumn{1}{c}{Expt.} \\
\hline
Si & VBW & 11.96 & 11.72 & 11.19 & 11.43 & 12.50\footnotemark[1]\footnotetext[1]{Ref.~\onlinecite{Faleev2004}} \\
\vspace{1mm} & 1CBW &  3.71 &  3.51 &  3.69 &  3.54 &  \\
Ge & VBW & 12.50 & 12.26 & 11.86 & 11.92 & 12.60\footnotemark[1] \\
\vspace{1mm} & 1CBW &  4.16 &  4.06 &  4.14 &  4.16 &  \\
SiO$_2$ & VBW$_{\mathrm{O}_{2p}}$ & 3.27 & 2.65 & 3.66 & 3.70 & 4.0\footnotemark[2]\footnotetext[2]{Ref.~\onlinecite{Gupta1985}} \\
        & VBW$_{\mathrm{O}_{2p}\,\mathrm{Si}_{3s,3p}}$ & 4.81 & 3.98 & 5.09 & 5.15 & 5.0\footnotemark[3]\footnotetext[3]{Ref.~\onlinecite{Laughlin1979}} \\
        & VBW$_{\mathrm{O}_{2s}}$ & 2.38 & 1.99 & 2.17 & 2.19 & 2.5\footnotemark[2] \\
\vspace{1mm} & 1CBW & 2.92 & 2.72 & 3.20 & 3.51 &  \\
ZnO & VBW & 6.19 & 5.41 & 6.43 & 6.49 & 9.0\footnotemark[4]\footnotetext[4]{Ref.~\onlinecite{Ozgur2005}} \\
\vspace{1mm} & 1CBW & 7.11 & 6.55 & 7.53 & 7.25 &  \\
SnO & VBW & 9.11 & 9.12 & 9.67 & 10.00 & 12.0 \footnotemark[5]\footnotetext[5]{Ref.~\onlinecite{Themlin1992}} \\
\vspace{1mm} & 1CBW & 2.98 & 3.22 & 3.35 &  3.60 &  \\
SnO$_2$ & VBW & 8.38 & 6.78 & 8.29 & 8.32 & 10.4\footnotemark[6]\footnotetext[6]{Ref.~\onlinecite{Ogo2008}} \\
\vspace{1mm} & 1CBW & 5.08 & 4.13 & 5.48 & 4.97 &  \\
CaS & VBW & 3.18 & 2.59 & 3.13 & 3.18 & 3.9\footnotemark[7] \footnotetext[7]{Ref.~\onlinecite{Chen2007} [Theor.]} \\
\vspace{1mm} & 1CBW & 3.09 & 3.07 & 3.43 & 4.28 &  \\
CaO & VBW$_{\mathrm{O}_{2p}}$ & 2.68 & 1.92 & 2.82 & 2.72 &  \\
    & VBW$_{\mathrm{O}_{2s}}$ & 1.22 & 0.86 & 0.88 & 1.42 &  \\
    & VBW$_{\mathrm{O}_{2p}}\Gamma$-X & 1.76 & 1.36 & 1.89 & 1.85 & 1.2\footnotemark[8]\footnotetext[8]{Ref.~\onlinecite{Bolorizadeh2004}} \\
    & VBW$_{\mathrm{O}_{2s}}\Gamma$-X & 0.55 & 0.37 & 0.40 & 0.76 & 0.6\footnotemark[8] \\
\vspace{1mm} & 1CBW & 3.47 & 2.55 & 3.72 & 4.36 &  \\
MgO & VBW & 4.71 & 3.71 & 5.05 & 5.01 & 4.8\footnotemark[9]\footnotetext[9]{Ref.~\onlinecite{Tjeng1990}} \\
\vspace{1mm} & 1CBW & 6.79 & 6.10 & 7.28 & 6.68 &  \\
LiF & VBW & 3.12 & 2.00 & 3.39 & 3.21 & 3.5\footnotemark[10]\footnotetext[10]{Ref.~\onlinecite{Himpsel1992}} \\
    & 1CBW & 5.84 & 5.64 & 6.24 & 6.21 & \\
\end{tabular}
\end{ruledtabular}
\end{table}


\subsection{Tuning the semi-empirical parameter of the TB09 functional}

Recently the authors of the TB09 proposed an improvement~\cite{Koller2012} of the original XC by redefining the parameters $\alpha$ and $\beta$ entering in the evaluation of the $c$ parameter of the functional using a larger set of materials. They claim that the TB09 and its revised version yield very accurate electronic band structures and gaps. According to the present study, the DFT valence band structures obtained with the TB09 are not guaranteed to reproduce the $G_0W_0$ band structures nor the experiment. In particular, for large band gap materials, the valence band widths are strongly contracted. Some studies are concerned with the determination of the best $c$ parameter which can always be tuned such that the experimental band gap is obtained. As already noted in the original TB09 publication, the band gap increases as the $c$ parameter is increased.

\begin{figure*}
\center
\hspace*{-0.1cm}
\includegraphics{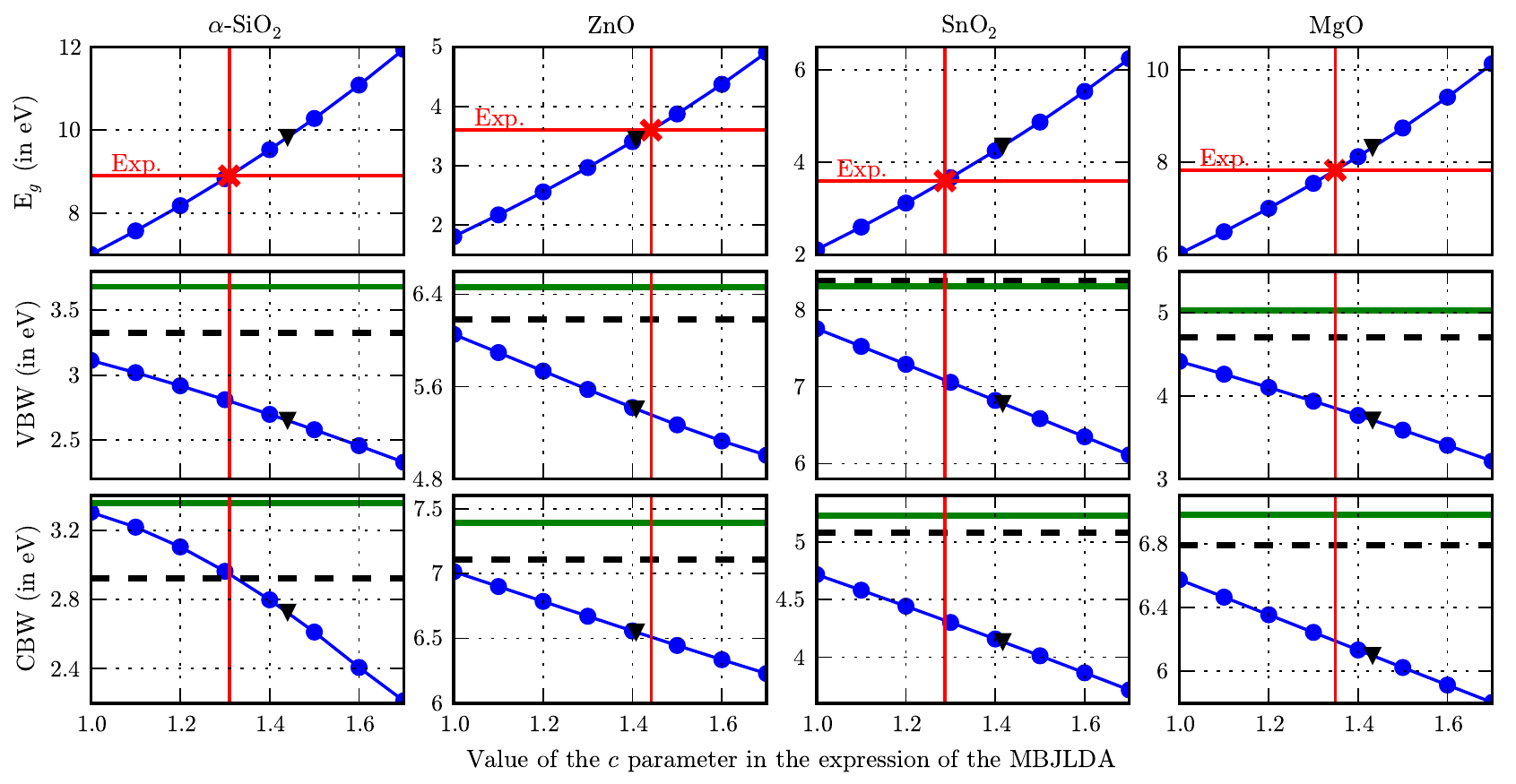}
\caption{\label{fig:c_param_gap_vbw}{(Color online) Fundamental band gaps E$_g$, upper valence band widths VBW and first conduction band widths 1CBW obtained by fixing the value of the $c$ parameter in Eq.~(\ref{eqn:vx_tb09}) for $\alpha$-SiO$_2$, ZnO, SnO$_2$ and MgO. The red crosses indicate the $c$ parameter that will reproduce the experimental band gap. The solid green horizontal lines in the VBW and 1CBW correspond to the $G_0W_0$ result. The black dashed horizontal lines correspond to the LDA value obtained in the present study. The black triangles are the actual results obtained with the TB09. It is clear that the $best$ $c$ value for the band gap will always narrow the band widths.}}
\vspace*{-0.5cm}
\end{figure*}

We have examined the possibility to tune the $c$ parameter for the materials in which the band width was clearly underestimated. Fig.~\ref{fig:c_param_gap_vbw} illustrates this procedure for $\alpha$-SiO$_2$, ZnO, SnO$_2$ and MgO. Similar trends are also observed for the other compounds. The modification of the upper valence band width and the conduction band is shown for various values of the $c$ parameter ranging from $1.0$ to $1.7$. It is clear that the band widths are decreased when the $c$ parameter is increased. This confirms our conjecture that the TB09 leads to band gaps larger than LDA by introducing a larger contribution of local exchange which at the same time contracts the electronic bands. So, no value of this $c$ parameter giving reasonable band gaps would correct the TB09's failure to describe the band widths correctly. The band widths obtained with LDA (black horizontal dashed lines) are always closer to the more reliable $G_0W_0$ results (green horizontal continuous lines) than those obtained with the TB09.

\section{\label{sec:conclusion}Conclusion}

A systematic investigation of electronic band structures obtained either with the LDA or the TB09 XC potentials has been performed for several oxides and other semiconducting and transparent materials. This recently proposed functional has attracted considerable attention as it is claimed to yield accurate electronic structure and band gaps. While the values of the band gaps obtained with the TB09 are indeed much improved over LDA, we have shown that the overall electronic structure can be unreliable especially for large band gap materials.

Comparisons with $G_0W_0$ results have been carried out for all the materials revealing that the TB09 can strongly underestimate the band widths. This contraction of the bands is found to be more pronounced for materials with a larger band gap. This follows from the fact that the TB09 opens the gaps by virtually adding a larger part of local exchange to the system through the adjustment of the $c$ parameter. This stronger exchange part results in a stronger localization of electron hence narrowing the band widths.

As a consequence, the ground-state electronic band structure is poorly described for wide band gap materials. This is expected to be detrimental for any further investigations such as defect studies, analysis of surfaces or interfaces, etc. It is shown that any modification of the $c$ parameter aiming to reproduce exactly the experimental band gap would not lead to any improvement. Indeed, the band widths of the materials are roughly proportional to the inverse of this parameter and will invariably be compressed.

Although very appealing for its efficiency and simplicity and for its performance in reproducing experimental band gaps of pure bulk materials, the TB09 must be used with great caution if any other properties than the band gap itself are under investigation. The perspectives of this kind of approaches are yet promising and there are possible avenues for the design of better XC functionals. In order to better reflect the actual many-body effects, a local XC functional should take into account some part of semi-locality such as what is done with the TB09 functional.

\section*{Acknowledgments}

The authors would like to thank Yann Pouillon, Alain Jacques and Jean-Michel Beuken for their valuable technical support and help with the build system of Abinit. D.~W. would like to acknowledge the support from the FNRS-FRIA. A.~L. and G.-M.R. acknowledge financial support from FRS-FNRS.
This work was supported by the Interuniversity
Attraction Poles program (P6/42) - Belgian State - Belgian
Science Policy, the IWT project Number 080023 (ISIMADE),
the Communaut\'{e} fran\c{c}aise de Belgique, through
the Action de Recherche Concert\'{e}e 07/12-003 Nanosyst\`{e}mes
hybrides m\'{e}tal-organiques", the FRS-FNRS through
FRFC project Number 2.4.589.09.F and 2.4645.08, the
R\'{e}gion Wallonne through WALL-ETSF project Number 816849.
M.J.T.O. would like to acknowledge useful discussions with M.A.L. Marques, financial support from the Portuguese FCT (contract \#SFRH/BPD/44608/2008), and FRS-FNRS support for a short stay in Louvain-la-Neuve.

\begin{thebibliography}{9}
\bibitem{Hohenberg1964}
P.~Hohenberg and W.~Kohn, Phys. Rev. {\bf 136}, B864 (1964).
\bibitem{Kohn1965}
W.~Kohn and L.~J.~Sham, Phys. Rev. {\bf 140}, A1133 (1965).
\bibitem{Burke2012}
K.~Burke, J. Chem. Phys. {\bf 136}, 150901 (2012).
\bibitem{Sham1985}
L. J. Sham and M. Schl\"uter, Phys. Rev. B {\bf 32}, 3883 (1985).
\bibitem{Hedin1965}
{L.}~Hedin, Phys. Rev. {\bf 139}, A796 (1965).
\bibitem{Hedin1970}
{L.}~Hedin and S.~Lundqvist, Solid State Phys. {\bf 23}, 1 (1970).
\bibitem{Aulbur2000}
W.~G.~Aulbur, L.~J\"onsson and J.~W.~Wilkins, \textit{Quasiparticle Calculations in Solids}, Solid State Phys. {\bf 54}, 1 (2000).
\bibitem{Gupta1985}
R. P. Gupta, Phys. Rev. B {\bf 32}, 8278 (1985).
\bibitem{Martin2004}
R. M. Martin, \textit{Electronic Structure : Basic Theory and Practical Methods}, (Cambridge University Press, Cambridge, England, 2004).
\bibitem{Chan2010}
M. K. Y. Chan and G. Ceder, Phys. Rev. Lett. {\bf 105}, 196403 (2010).
\bibitem{Marques2011}
M.~A.~L.~Marques, J.~Vidal, M.~J.~T.~Oliveira, L.~Reining and S.~Botti, Phys. Rev. B {\bf 83}, 035119 (2011).
\bibitem{Tran2009}
F.~Tran and P.~Blaha, Phys. Rev. Lett. {\bf 102}, 226401 (2009).
\bibitem{Becke2006}
A.~D. Becke and E.~R. Johnson, J. Chem. Phys. {\bf 124}, 221101 (2006).
\bibitem{Camargo2012}
J.~A. Camargo-Mart\'{i}nez, and R. Baquero, Phys. Rev. B {\bf 86}, 195106 (2012).
\bibitem{Kresse2012}
G. Kresse, M. Marsman, L. E. Hintzsche, and E. Flage-Larsen, Phys. Rev. B {\bf 85}, 045205 (2012).
\bibitem{Smith2012}
P. V. Smith, M. Hermanowicz, G. A. Shah and M. W. Radny, Comp. Mater. Sci. {\bf 54}, 37 (2012).
\bibitem{Feng2010}
W. Feng, D. Xiao, Y. Zhang and Y. Yao, Phys. Rev. B {\bf 82}, 235121 (2010).
\bibitem{Dixit2012}
H. Dixit, R. Saniz, S. Cottenier, D. Lamoen and B. Partoens, J. Phys.: Condens. Matter {\bf 24}, 205503 (2012).
\bibitem{Hetaba2012}
W. Hetaba, P. Blaha, F. Tran and P. Schattschneider, Phys. Rev. B {\bf 85}, 205108 (2012).
\bibitem{Singh2010}
D. J. Singh, Phys. Rev. B {\bf 82}, 205102 (2010).
\bibitem{Karolewski2009}
A. Karolewski, R. Armiento, and S. K\"ummel, J. Chem. Theory Comput. {\bf 5}, 712 (2009).
\bibitem{Gaiduk2009}
A. P. Gaiduk and V. N. Staroverov, J. Chem. Phys. {\bf 131}, 044107 (2009).
\bibitem{Koller2011}
D. Koller, F. Tran and P. Blaha, Phys. Rev. B {\bf 83}, 195134 (2011).
\bibitem{Kim2010}
Y.-S. Kim, M. Marsman, G. Kresse, F. Tran and P. Blaha, Phys. Rev. B {\bf 82}, 205212 (2010).
\bibitem{Raesaenen2010}
E.~R\"as\"anen, S.~Pittalis and C.~R.~Proetto, J. Chem. Phys. {\bf 132}, 044112 (2010).
\bibitem{xps_broadening}
The experimental XPS spectra differ from the theoretical DOS for various reasons. Depending on the conditions in which the experiments were performed, on the temperature, on the quality of the factors and other factors such as surface effects, the XPS data are broadened with respect to the theoretical results. Note that the DOS and XPS spectra also differ by the matrix elements.  We refer the reader to Ref.~\onlinecite{Guzzo2011} for a detailed discussion of these aspects.
\bibitem{Guzzo2011}
M.~Guzzo, G.~Lani, F.~Sottile, P.~Romaniello, M.~Gatti, J.~J.~Kas, J.~J.~Rehr, M.~G.~Silly, F.~Sirotti and L.~Reining, Phys. Rev. Lett. {\bf 107}, 166401 (2011).
\bibitem{Gonze2005}
X. Gonze \textit{et al.}, Z. Kristallogr. {\bf 220}, 558 (2005).
\bibitem{Gonze2009}
X. Gonze \textit{et al.}, Comput. Phys. Commun. {\bf 180}, 2582 (2009).
\bibitem{Oliveira2008}
M.~J.~Oliveira and F.~Nogueira, Comput. Phys. Commun. {\bf 178}, 524 (2008).
\bibitem{Marques2012}
M.~A.~Marques, M.~J.~Oliveira and T.~Burnus, Comput. Phys. Commun. {\bf 183}, 2272 (2012).
\bibitem{Perdew1992a}
J.~P. Perdew and Y.~Wang, Phys. Rev. B {\bf 45}, 13244 (1992).
\bibitem{Ceperley1980}
D.~M. Ceperley and B.~J. Alder, Phys. Rev. Lett. {\bf 45}, 566 (1980).
\bibitem{Becke1989}
A.~D. Becke and M.~R. Roussel, Phys. Rev. A {\bf 39}, 3761 (1989).
\bibitem{Godby1989}
R.~W. Godby and R.~J. Needs, Phys. Rev. Lett. {\bf 62}, 1169 (1989).
\bibitem{Stankovski2011}
M. Stankovski, G. Antonius, D. Waroquiers, A. Miglio, H. Dixit, K. Sankaran, M. Giantomassi, X. Gonze, M. C\^ot\'e, and G.-M. Rignanese, Phys. Rev. B {\bf 84}, 241201 (2011).
\bibitem{Miglio2012}
A. Miglio, D. Waroquiers, G. Antonius, M. Giantomassi, M. Stankovski, M. C\^ot\'e, X. Gonze, and G.-M. Rignanese,
Eur. Phys. J. B {\bf 85}, 322 (2012).
\bibitem{Bruneval2008}
F. Bruneval and X. Gonze, Phys. Rev. B {\bf 78}, 085125 (2008).
\bibitem{Schilfgaarde2006}
M. van Schilfgaarde, T. Kotani and S. Faleev, Phys. Rev. Lett. {\bf 96}, 226402 (2006).
\bibitem{Marzari1997}
N. Marzari and D. Vanderbilt, Phys. Rev. B {\bf 56}, 12847 (1997).
\bibitem{Souza2001}
I. Souza, N. Marzari and D. Vanderbilt, Phys. Rev. B {\bf 65}, 035109 (2001).
\bibitem{Shaltaf2009}
R. Shaltaf, T. Rangel, M. Gr\"uning, X. Gonze, G.-M. Rignanese, and D.R. Hamann, Phys. Rev. B {\bf 79}, 195101 (2009).
\bibitem{Vidal2010a}
J. Vidal, F. Trani, F. Bruneval, M.~A.~L. Marques and S. Botti, Phys. Rev. Lett. {\bf 104}, 136401 (2010).
\bibitem{Wyckoff1960}
R.~W.~G.~Wyckoff, \textit{Crystal structures} (John Wiley \& Sons, New York, London, 1963).
\bibitem{Ley1972}
L. Ley, S. Kowalczyk, R. Pollak, and D. A. Shirley, Phys. Rev. Lett. {\bf 29}, 1088 (1972).
\bibitem{Eastman1974}
D. E. Eastman, W. D. Grobman, J. L. Freeouf and M. Erbudak, Phys. Rev. B {\bf 9}, 3473 (1974).
\bibitem{Kasper2005}
E. Kasper and D. Paul, \textit{Silicon Quantum Integrated Circuits. Silicon-germanium Heterostructure Devices : Basics and Realisation} (Springer, Berlin, 2005).
\bibitem{Faleev2004}
S. V. Faleev, M. van Schilfgaarde, and T. Kotani, Phys. Rev. Lett. {\bf 93}, 126406 (2004).
\bibitem{Laughlin1980}
R. B. Laughlin, Phys. Rev. B {\bf 22}, 3021 (1980).
\bibitem{Weinberg1979}
Z. A. Weinberg, G. W. Rubloff, and E. Bassous, Phys. Rev. B {\bf 19}, 3107 (1979).
\bibitem{Wyckoff1963}
R. W. G. Wyckoff, \textit{Crystal Structures}, Vol. 1 (John Wiley \& Sons, New York, London, 1963).
\bibitem{Zakaznova-Herzog2005}
V. P. Zakaznova-Herzog, H. W. Nesbitt, G. M. Bancroft, J. S. Tse, X. Gao, and W. Skinner, Phys. Rev. B {\bf 72}, 205113 (2005).
\bibitem{Chang2000}
E. K. Chang, M. Rohlfing and S. G. Louie, Phys. Rev. Lett. {\bf 85}, 2613 (2000).
\bibitem{Laughlin1979}
R. B. Laughlin, J. D. Joannopoulos and D. J. Chadi, Phys. Rev. B {\bf 20}, 5228 (1979).
\bibitem{Kihara1985}
K. Kihara and G. Donnay, Can. Mineral. {\bf 23}, 647 (1985).
\bibitem{King2009}
P. D. C. King, T. D. Veal, A. Schleife, J. Z\'{u}\~{n}iga-P\'{e}rez, B. Martel, P. H. Jefferson, F. Fuchs, V. Mu\~{n}oz-Sanjos\'{e}, F. Bechstedt and C. F. McConville, Phys. Rev. B {\bf 79}, 205205 (2009).
\bibitem{Lany2005}
S. Lany and A. Zunger, Phys. Rev. B {\bf 72}, 035215 (2005).
\bibitem{Laskowski2006}
R. Laskowski and N. E. Christensen, Phys. Rev. B {\bf 73}, 045201 (2006).
\bibitem{Lany2007}
S. Lany and A. Zunger, Phys. Rev. Lett. {\bf 98}, 045501 (2007).
\bibitem{Zhou2008}
X. H. Zhou, Q.-H. Hu and Y. Fu, J. Appl. Phys. {\bf 104}, 063703 (2008).
\bibitem{Vogel1995}
D. Vogel, P. Kr\"{u}ger, and J. Pollmann, Phys. Rev. B {\bf 52}, R14316 (1995). 
\bibitem{Ozgur2005}
\"{U}. \"{O}zg\"{u}r, Y. I. Alivov, C. Liu, A. Teke, M. A. Reshchikov, S. Dogan, V. Avrutin, S.-J. Cho, and H. Morko\c{c}, J. Appl. Phys. {\bf 98}, 041301 (2005).
\bibitem{Tsoi2006}
S. Tsoi, X. Lu, A. K. Ramdas, H. Alawadhi, M. Grimsditch, M. Cardona and R. Lauck, Phys. Rev. B {\bf 74}, 165203 (2006).
\bibitem{Pannetier1980}
J. Pannetier and G. Denes, Acta Cryst. B {\bf 36}, 2763 (1980).
\bibitem{Haines1997}
J. Haines and J. M. L\'eger, Phys. Rev. B {\bf 55}, 11144 (1997).
\bibitem{Themlin1992}
J.-M. Themlin, M.~Chta\"ib, L.~Henrard, P.~Lambin, J.~Darville and J.-M.~Gilles, Phys. Rev. B {\bf 46}, 2460 (1992).
\bibitem{Bruneval2006}
F. Bruneval, N. Vast and L. Reining, Phys. Rev. B {\bf 74}, 045102 (2006).
\bibitem{Ogo2008}
Y. Ogo, H. Hiramatsu, K. Nomura, H. Yanagi, T. Kamiya, M. Hirano and H. Hosono, Appl. Phys. Lett. {\bf 93}, 032113 (2008).
\bibitem{Batzill2005}
M. Batzill and U. Diebold, Prog. Surf. Sci. {\bf 79}, 47 (2005).
\bibitem{Kaneko1990}
Y. Kaneko and T. Koda, J. Cryst. Growth {\bf 86}, 72 (1990).
\bibitem{Chen2007}
Z. Chen, H. Xiao and X. Zu, Physica B {\bf 391}, 193 (2007).
\bibitem{Smith1968}
D. K. Smith and H. R. Leider, J. Appl. Cryst. {\bf 1}, 246 (1968).
\bibitem{Whited1969}
R. C. Whited and W. C. Walker, Phys. Rev. Lett. {\bf 22}, 1428 (1969).
\bibitem{Whited1973}
R. C. Whited, C. J. Flaten and W. C. Walker, Solid State Commun. {\bf 13}, 1903 (1973).
\bibitem{Bolorizadeh2004}
M.A. Bolorizadeh, V.A. Sashin, A.S. Kheifets and M.J. Ford, J. Electron. Spectrosc. Relat. Phenom. {\bf 141}, 27 (2004).
\bibitem{Bobade2012}
S. M. Bobade, Appl. Phys. Lett. {\bf 100}, 072102 (2012).
\bibitem{French1990}
R.H. French, R. V. Kasowski, F. S. Ohuchi, D. J. Jones, H. Song and R. L. Coble, J. Am. Ceram. Soc. {\bf 73}, 3195 (1990).
\bibitem{Tjeng1990}
L. Tjeng, A. Vos and G. Sawatzky, Surf. Sci. {\bf 235}, 269 (1990).
\bibitem{Ott1926}
H. Ott, Z. Kristallogr. Krist. {\bf 63}, 222 (1926).
\bibitem{Himpsel1992}
F.~J.~Himpsel, L.~J.~Terminello, D.~A.~Lapiano-Smith, E.~A.~Eklund and J.~J.~Barton, Phys. Rev. Lett. {\bf 68}, 3611 (1992).
\bibitem{Piacentini1976}
M.~Piacentini, D.~W.~Lynch and C.~G.~Olson, Phys. Rev. B {\bf 13}, 5530 (1976).
\bibitem{Shishkin2007a}
M. Shishkin, M. Marsman and G. Kresse, Phys. Rev. Lett. {\bf 99}, 246403 (2007).
\bibitem{Koller2012}
D. Koller, F. Tran and P. Blaha, Phys. Rev. B {\bf 85}, 155109 (2012).
\end{thebibliography}

\end{document}